\documentclass[sigconf, preprint]{acmart}

\AtBeginDocument{%
  \providecommand\BibTeX{{%
    \normalfont B\kern-0.5em{\scshape i\kern-0.25em b}\kern-0.8em\TeX}}}

\setcopyright{acmcopyright}
\copyrightyear{2018}
\acmYear{2018}
\acmDOI{XXXXXXX.XXXXXXX}

\acmConference[Conference acronym 'XX]{Make sure to enter the correct
  conference title from your rights confirmation emai}{June 03--05,
  2018}{Woodstock, NY}
\acmPrice{15.00}
\acmISBN{978-1-4503-XXXX-X/18/06}

\usepackage{multirow}
\usepackage{graphicx}

\newcommand{\eg}{\emph{e.g., }}

\newcommand{\zjz}[1]{{#1}}
\newcommand{\zjzz}[1]{{#1}}

\newcommand{\swt}[1]{{#1}}

\newcommand{\aka}

\usepackage{enumitem}
\usepackage{verbatim}
\usepackage[ruled]{algorithm2e}
\usepackage{multirow}
\usepackage{subfigure} 
\usepackage{ragged2e}
\usepackage{caption}
\usepackage{graphicx}
\usepackage[normalem]{ulem}
\useunder{\uline}{\ul}{}
\usepackage{amsmath}
\usepackage{setspace}
\usepackage{bm}
\usepackage{longtable}
\usepackage{mathrsfs}
\newcommand{\name}{Rec4Agentverse}
\newcommand{\recname}{Agent Recommender}
\newcommand{\itemname}{Agent Item}

\begin{document}



\title{\zjzz{Prospect Personalized Recommendation on Large Language Model-based Agent Platform}}


\settopmatter{authorsperrow=3}
\author{Jizhi Zhang*\dag}
\email{cdzhangjizhi@mail.ustc.edu.cn}
\affiliation{%
  \institution{University of Science and Technology of China}
  \city{}
  \country{}
  }

\author{Keqin Bao*}
\email{baokq@mail.ustc.edu.cn}
\affiliation{%
  \institution{University of Science and Technology of China}
  \city{}
  \country{}
  }

\author{Wenjie Wang}
\email{wangwenjie@u.nus.edu}
\affiliation{%
  \institution{National University of Singapore}
  \city{}
  \country{}
}

\author{Yang Zhang}
\email{zy2015@mail.ustc.edu.cn}
\affiliation{%
  \institution{University of Science and Technology of China}
  \city{}
  \country{}
}

\author{Wentao Shi}
\email{shiwentao123@mail.ustc.edu.cn}
\affiliation{%
  \institution{University of Science and Technology of China}
  \city{}
  \country{}
}

\author{Wanhong Xu}
\email{wanhong.xu@gmail.com}
\affiliation{%
  \institution{Yuanshi Technology Inc. }
  \city{}
  \country{}
}

\author{Fuli Feng}
\email{fulifeng93@gmail.com}
\affiliation{%
  \institution{University of Science and Technology of China}
  \city{}
  \country{}
}

\author{Tat-Seng Chua}
\email{dcscts@nus.edu.sg}
\affiliation{%
  \institution{National University of Singapore}
  \city{}
  \country{}
}

\thanks{*Equal Contribution. \\ $\dagger$ 
This work is done when Jizhi Zhang is a research intern at Yuanshi.
}


\renewcommand{\shortauthors}{Jizhi Zhang, et al.}

\begin{abstract}

\zjzz{The new kind of Agent-oriented information system, exemplified by GPTs, urges us to inspect the information system infrastructure to support Agent-level information processing and to adapt to the characteristics of Large Language Model (LLM)-based Agents, such as interactivity. 
In this work, we envisage the prospect of the recommender system on LLM-based Agent platforms and introduce a novel recommendation paradigm called Rec4Agentverse, comprised of Agent Items and Agent Recommender.
Rec4Agentverse emphasizes the collaboration between Agent Items and Agent Recommender, 
thereby promoting personalized information services and enhancing the exchange of information beyond the traditional user-recommender feedback loop. 
Additionally, we prospect the evolution of Rec4Agentverse and conceptualize it into three stages based on the enhancement of the interaction and information exchange among Agent Items, Agent Recommender, and the user.
A preliminary study involving several cases of Rec4Agentverse validates its significant potential for application. 
Lastly, we discuss potential issues and promising directions for future research.
}

\end{abstract}



\settopmatter{printfolios=true}
\maketitle
\section{Introduction}

\begin{figure}[htbp]
\centering
\vspace{+10pt}
\includegraphics[width=0.48\textwidth]{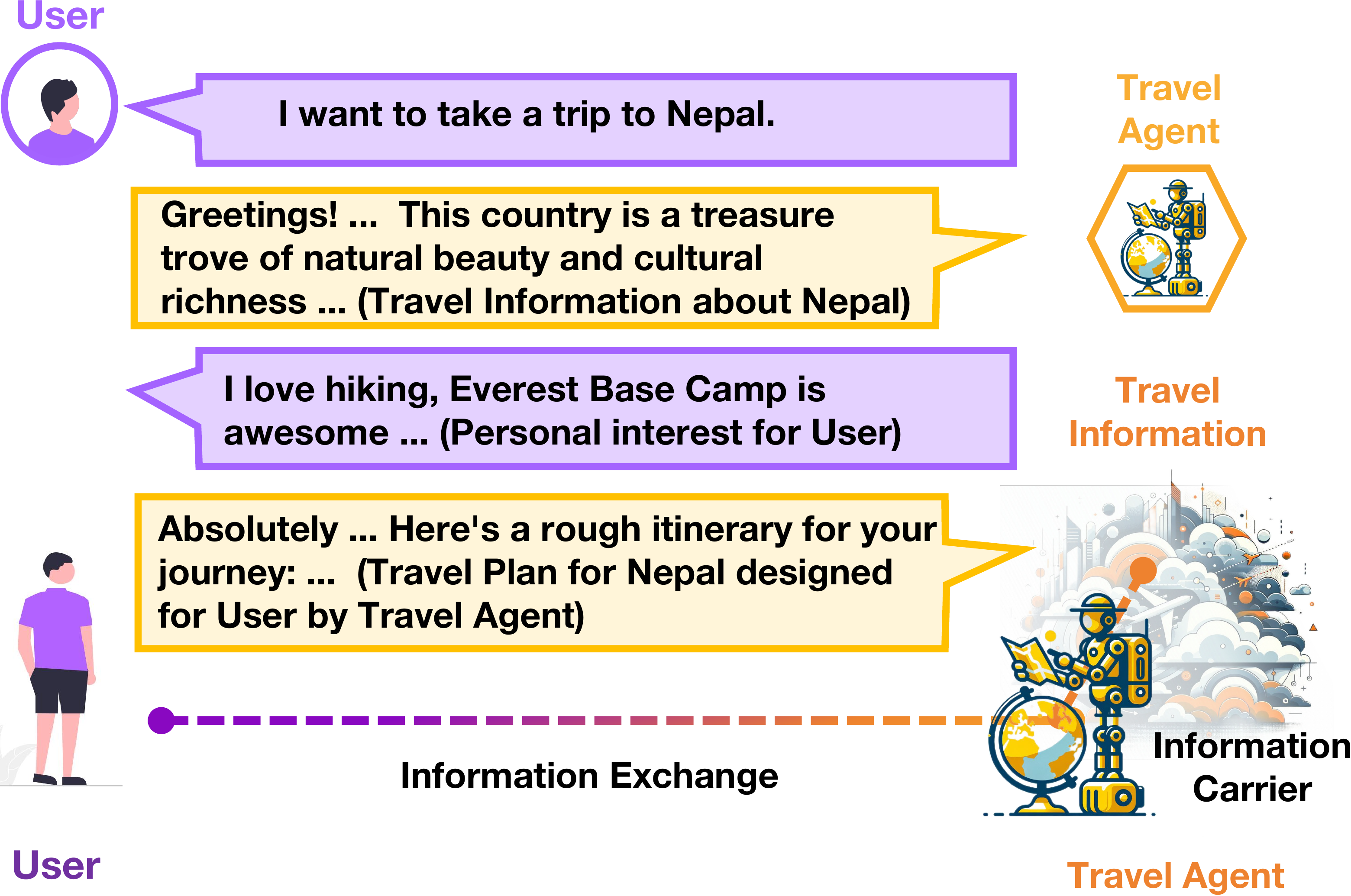}
\caption{An example of interaction between an ~\itemname~ and a user.
A Travel Agent can serve as an information carrier with travel-related information, as well as engage in a dialogue with the user to exchange related information. 
}
\label{fig:intro_case} 
\end{figure}

\begin{figure*}[htbp]
\centering
\includegraphics[width=\textwidth]{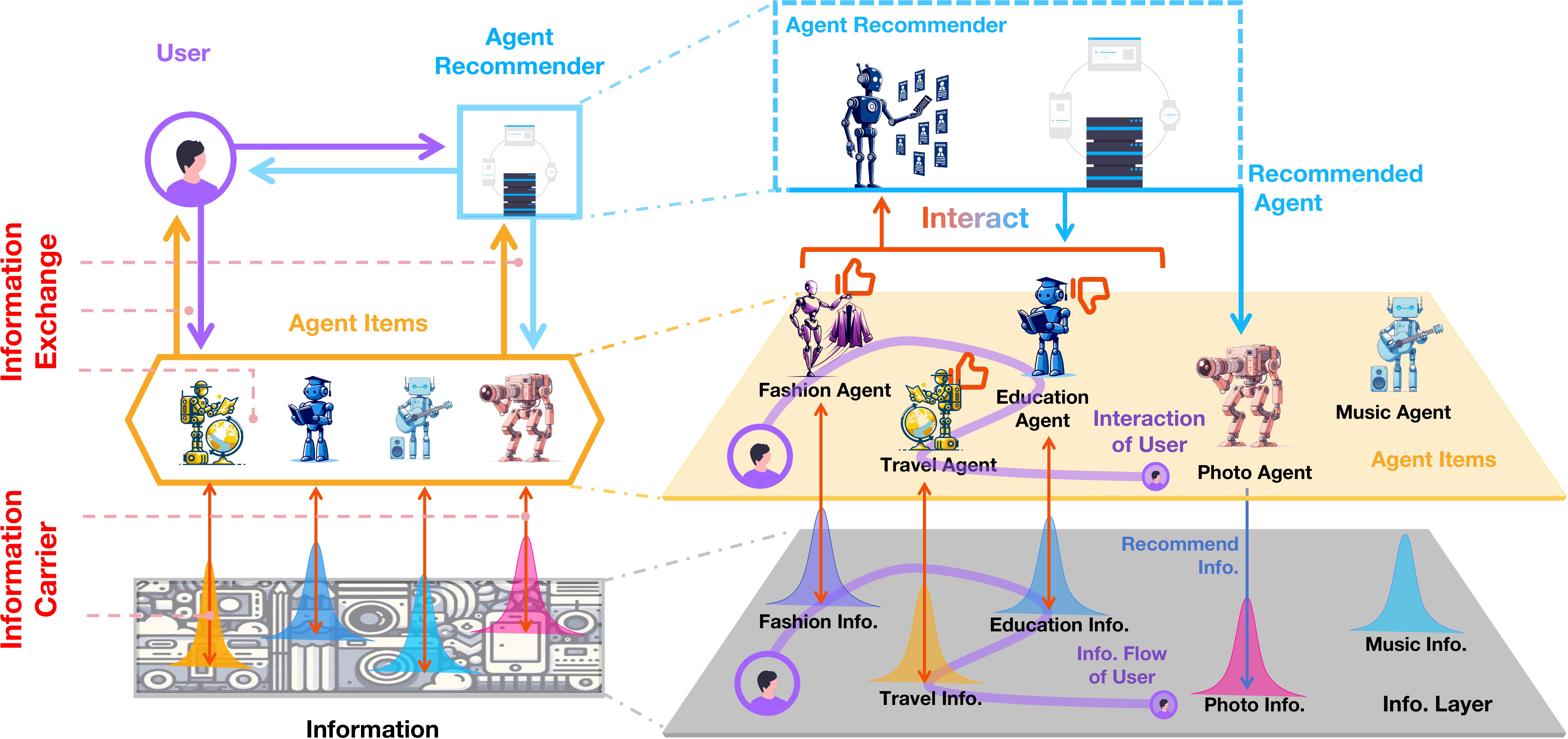}
\caption{Illustration of the \name~paradigm. 
The left portion of the diagram depicts three roles in RecAgentverse: user, Agent Recommender, and Agent Item, along with their interconnected relationships. 
In contrast to traditional recommender systems, Rec4Agentverse has more intimate relationships among the three roles. 
For instance, there are multi-round interactions between 1) users and Agent Items, and 2) Agent Recommender and Agent Items. 
The right side of the diagram demonstrates that Agent Recommender can collaborate with Agent Items to affect the information flow of users and offer personalized information services. 
} 
\label{fig:overview} 
\end{figure*}

Large Language Model (LLM)-based Agents have garnered widespread attention in various fields due to their astonishing capabilities such as natural language communication~\cite{park2023generative, lin2023llm}, instruction following~\cite{agent_survey_1, song2023llm}, and task execution abilities~\cite{plan_1, task_1, zhao2023survey}. 
\zjz{Such astonishing capabilities hold the potential to extend the format of information carriers and the way of information exchange.}
On one hand, 
LLM-based Agents can evolve into various domain experts, forming novel information carriers with domain-specific knowledge~\cite{diverse_agent_1, agent_survey_1}. 
\zjz{For example, a Travel Agent can retain travel-related information within its parameters.}
On the other hand, LLM-based Agents showcase a novel form of information exchange, facilitating more intuitive and natural interactions with users through dialogue and task execution~\cite{agent_survey_2, shao2023character}.
Figure~\ref{fig:intro_case} shows an example of such information exchange, where users engage in dialogue with a Travel Agent to obtain travel information and complete travel planning.




Along with the increase of LLM-based Agents in various domains, Agent platforms (e.g., GPTs\footnote{\url{https://chat.openai.com/gpts}.}) represent a novel kind of information system with Agent-oriented information gathering, storing, and exchanging.
\zjz{Consequently, the infrastructure of information systems needs to be expanded to support information processing at the Agent level and accommodate the significant properties of Agents like interactivity, intelligence, and proactiveness~\cite{agent_survey_1, agent_survey_2}.}
\zjz{Within the infrastructure, the recommender system is a key cornerstone, which greatly affects how information flows in the information system regarding efficiency, user experience, and many other factors. Therefore, it is essential to envision how the recommender system can function on the LLM-based Agent platform.
}
\zjz{To this end, we propose a novel recommendation paradigm for the LLM-based Agent platform, named Rec4Agentverse. As illustrated in Figure~\ref{fig:overview}, Rec4Agentverse includes two key concepts: Agent Recommender and Agent Item.}
\zjz{Agent Item means treating LLM-based Agents as items in the recommender system. Agent Recommender is employed to recommend personalized Agent Items for each user.}
\zjz{In contrast to items in traditional recommender systems, Agent Items have properties such as interactivity, intelligence, and proactiveness. Such properties make it possible for Agent Items and Agent Recommender to collaborate and share user information\footnote{
It may involve user's privacy and requires user permission.
}, facilitating personalized information delivery.}
For example, once a Travel Agent is recommended to a 
user, it can continuously discern user's preferences regarding travel during their interaction and convey this preference back to the \recname. 

We envision three stages for the development of Rec4Agentverse to increasingly support the interaction and information exchange among the user, Agent Recommender, and Agent Items.
\begin{itemize}[leftmargin=*]
    \item \textbf{Stage 1: User-Agent Interaction}. 
    Rec4Agentverse adds the information exchange between the user and Agent Item.
    \zjz{For instance, Agent Recommender will recommend an Agent Item to a user based on personal needs and preferences. 
    Agent Item engages in a dialogue with the user, subsequently providing information for the user and also acquiring user information.}
\item \textbf{Stage 2: Agent-Recommender Collaboration}. 
Rec4Agentverse then enables the information exchange between Agent Item and Agent Recommender.
\zjz{For example, Agent Item can transmit the latest preferences of the user back to Agent Recommender. Agent Recommender can give new instructions to Agent Item.} 

\item \textbf{Stage 3: Agents Collaboration}. 
Rec4Agentverse then supports the collaboration among Agent Items to further facilitate the exchange of information and enhance the personalized information service of users. 
During this stage, diverse \itemname s can participate in information sharing and collaboration.
\end{itemize}

\zjz{We explore the preliminary instantiation of the Rec4Agentverse paradigm in some cases, showcasing its significant application potential.}
Furthermore, we introduce the potential application scenarios of \name, as well as the issues and challenges in the application, inspiring future exploration. 
Our contributions can be summarized as follows:
\begin{itemize}[leftmargin=*]
    \item We propose Rec4Agentverse, a novel recommendation paradigm for the LLM-based Agent platform, providing users with personalized agent services. 
    \item We envision three milestones in the evolution of \name. Besides, we introduce potential research directions, application domains, and application challenges of \name, significantly facilitating future exploration. 
    \item  We conduct the preliminary feasibility study for \name, demonstrating that existing LLM-based Agents are promising to instantiate \name. 
\end{itemize}

\section{\name~Paradigm}


%

In this section, we will give an overview of \name.
The LLM-based Agent platform emerges as a new information system in terms of novel information carriers and new ways of information exchange. We thus propose to revolutionize recommendation systems for this new information system and introduce \name. 
Firstly, we shall elucidate the different parts within \name~ (Section~\ref{sec:2.1}). 
Subsequently, we will contemplate the three stages of \name~from the perspective of information flow (Section~\ref{sec:2.2}). 
Lastly, we shall proffer potential applications of this paradigm in various domains (Section~\ref{sec:2.3}), explore pertinent research topics (Section~\ref{sec:2.4}), and discuss potential challenges and risks in the application (Section~\ref{sec:2.5}).
\subsection{Roles of \name}
\label{sec:2.1}
The \name~paradigm consists of three roles: the user, the ~\recname, and the ~\itemname as illustrated in Figure~\ref{fig:three_stage_intro}. 
The user, just like in traditional recommender systems, interacts with both Agetn Items and Agent Recommender and gives feedback. 
Therefore, our primary focus will be on discussing concepts that differ significantly from traditional recommendation systems, namely ~\itemname~ and ~\recname.

\zjz{
\subsubsection{\textbf{~\itemname}}
The ~\itemname~ is the most distinct aspect in the \name~paradigm compared to traditional recommendation paradigms. 
Unlike the conventional item in the traditional recommendation system, the item in the \name~paradigm transforms into an LLM-based Agent. 
As illustrated in Figure~\ref{fig:three_stage_intro}, the ~\itemname~ cannot only interact with users but also collaborate with the ~\recname~ and other ~\itemname s.
\zjzz{The creation process and origins of Agent Items could be diverse and varied. For instance, the creation process of Agent Items can involve training with domain-specific data or directly constructing Agent Items through prompts. The origin of Agent Item could be either generated automatically by the LLM-based Agent platform, created by users or collaboratively created by both users and the platform.}
}
\subsubsection{\textbf{~\recname}} 
\recname~ aims to recommend LLM-based agents to users. 
Its function is similar to that of traditional recommender systems, which infer user preferences based on collected user information (\eg attributes and behaviors) to recommend new items. 
However, unlike traditional recommender systems, the recommended items in ~\recname~ are LLM-based Agents, which imbues it with distinctive characteristics set apart from conventional recommenders. 

\begin{figure}[t]
\centering
\includegraphics[width=0.45\textwidth]{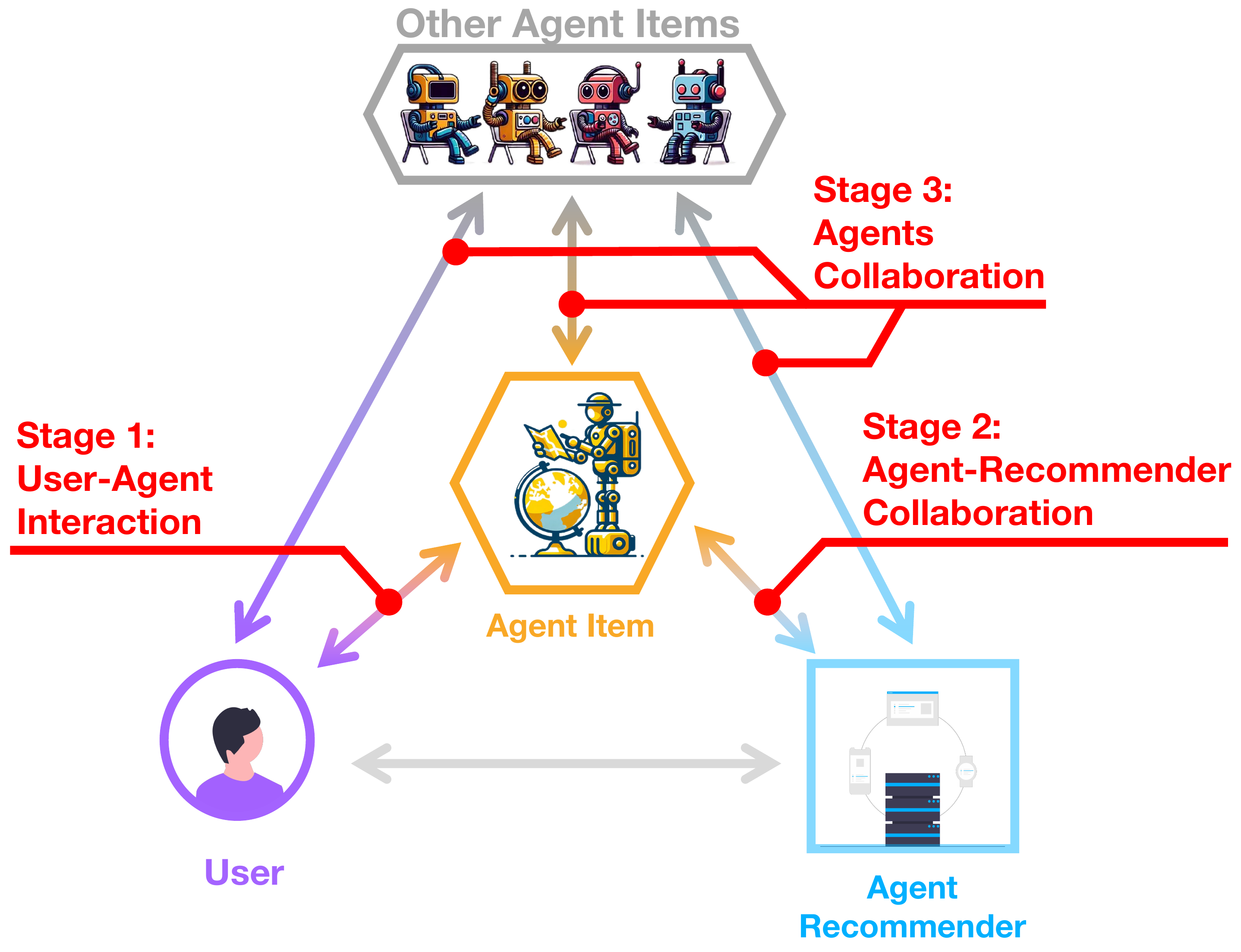}
\caption{Three stages of \name.
The bidirectional arrows depicted in the Figure symbolize the flow of information. During the first stage of User-Agent interaction, information flows between the user and Agent Item. In the Agent-Recommender Collaboration stage, information flows between Agent Item and Agent Recommender. For the Agents Collaboration stage, information flows between various Agent Items.
} 
\label{fig:three_stage_intro} 
\end{figure}

\zjz{Within the Rec4Agentverse paradigm, Agent Recommender is expected to possess enhanced capabilities for information exchange and collaboration with other parts of the Rec4Agentverse}. 
As illustrated in Figure~\ref{fig:three_stage_intro}, in this new paradigm, the ~\recname~ not only engages in direct interactions with users but also interacts with the ~\itemname, issuing commands to the ~\itemname~ or obtaining new feedback from users via ~\itemname.

\subsection{Three Stages of \name}
\label{sec:2.2}
In this subsection, we will discuss three key stages of our proposed \name~paradigm from the information flow perspective as illustrated in Figure~\ref{fig:three_stage_intro}.
In addition to the interaction between users and recommender systems in traditional recommender systems, \name~also takes into account the profound interaction between users and ~\itemname, as well as the collaboration between ~\itemname~ and ~\recname, and the collaboration between ~\itemname~ themselves. 
This formulation encompasses three collaboration scenarios, envisioning the future development path of ~\name. 
\subsubsection{\textbf{Stage 1: User-Agent Interaction}}
During the initial stage, in addition to the interaction between the user and the ~\recname, the user also engages in interaction with ~\itemname. 
This interactive format is similar to traditional recommendations. 
On LLM-based agent platforms such as GPTs, ~\name~ may generate or retrieve personalized LLM-based Agents according to explicit user instructions and implicit user behaviors. 
While users can interact with the LLM-based Agent to exchange information in a novel form, it does not fully unleash the immense potential of the LLM-based Agent. 
Aside from interacting with users, the ~\itemname~ can also collaborate with other roles in the recommender system to further enrich the information flow on the LLM-based Agent platform. 

\subsubsection{\textbf{Stage 2: Agent-Recommender Collaboration}}
In this stage, ~\itemname~ will collaborate with Agent Recommender together to provide information service for users. 
Different from items in the traditional recommender system, ~\itemname~ can deeply collaborate with ~\recname~ by feeding forward to and receiving user information from the ~\recname. 
\zjz{
For example, Agent Item can share the user preferences it collects with Agent Recommender so that Agent Recommender can better provide more personalized recommendations.
Similarly, Agent Items can also receive new instructions from Agent Recommender. 
The collected personalized information from users and instructions from Agent Recommender can be used to update Agent Item for evolvement (\eg prompt updates) so that Agent Item can better understand user preferences and provide superior information services. 
}

\subsubsection{\textbf{Stage 3: Agents Collaboration}}
An ~\itemname~ can collaborate with other Agent Items with different domain knowledge to provide diverse information services for users.
\zjz{A simple example is when 
a user mentions 
some niche things that Agent Item does not know about. Agent Item can put forward a request to Agent Recommender to ask Agent Recommender to recommend a new Agent Item for its assistance.
Then the two agents can collaborate to fulfill users' information needs or execute tasks. 
Beyond that, there is considerable room for imagination at this stage. For example, the recommended new Agent Item can also interact with users directly or with Agent Recommender. Further, if multiple Agent Items are recommended, these Agent Items can also work together to better complete the user's instructions through brainstorming or round-table meetings.
}

\subsection{Application Domains}
\label{sec:2.3}
Our \name~paradigm can contain Agent Items from various domains, which could originate from various third-party client developers or expert agents directly created by \recname.
\name~ can be applied to many scenarios, and here we provide a few illustrative examples in representative domains, showcasing the potentiality of our \name~framework.

\begin{itemize}[leftmargin = *]
    \item \textbf{Travel Agents}. 
    Travel Agents are designed to assist users in planning and booking travel arrangements. 
    \swt{When a user indicates a specific travel destination of interest, Agent Recommender can recommend a proficient Travel Agent who possesses expertise in facilitating travel arrangements. Subsequently, users can engage in interactions with the recommended Travel Agent to obtain personalized travel itineraries and plans.}
    The Travel Agent can further collect user information, either through direct interaction with users or by accessing the \recname, to infer users' individual preferences and upgrade itself for better travel recommendations. 
    \swt{Additionally, the Travel Agent can engage in collaborative efforts with other agents, thereby acquiring valuable insights regarding users' preferences from diverse domains. This collaborative approach enables the Travel Agent to offer users more adaptable and personalized travel plans.}
    \item \textbf{Fashion Agents}.
    Fashion Agents aim to assist users in discovering their preferred fashion styles and recommend fashion items that align with their preferences. 
    Similar to Travel Agents, Fashion Agents can engage in conversations with users or interact with Agent Recommender to gather users' fashion preferences. Agent Recommender may summarize user preferences through previously recommended Agent Items.  
    For example, Fashion Agents might gather user preferences about the places and local characteristics from the user interactions with a Travel Agent. 
    Moreover, a Fashion Agent can collaborate with a Tailor Agent to design and make personalized new clothes for a user. 

    \item \textbf{Sports Agents}.
    Sports Agent aims to recommend suitable exercise plans to users. 
    They can engage with users, \recname, and other \itemname s to collect user preferences and offer exercise plans and recommendations. 
    For example, they can use information about a user's physical condition obtained from Travel Agents to create suitable exercise plans.
\end{itemize}

\begin{figure*}[htbp]
\centering
\includegraphics[width=\textwidth]{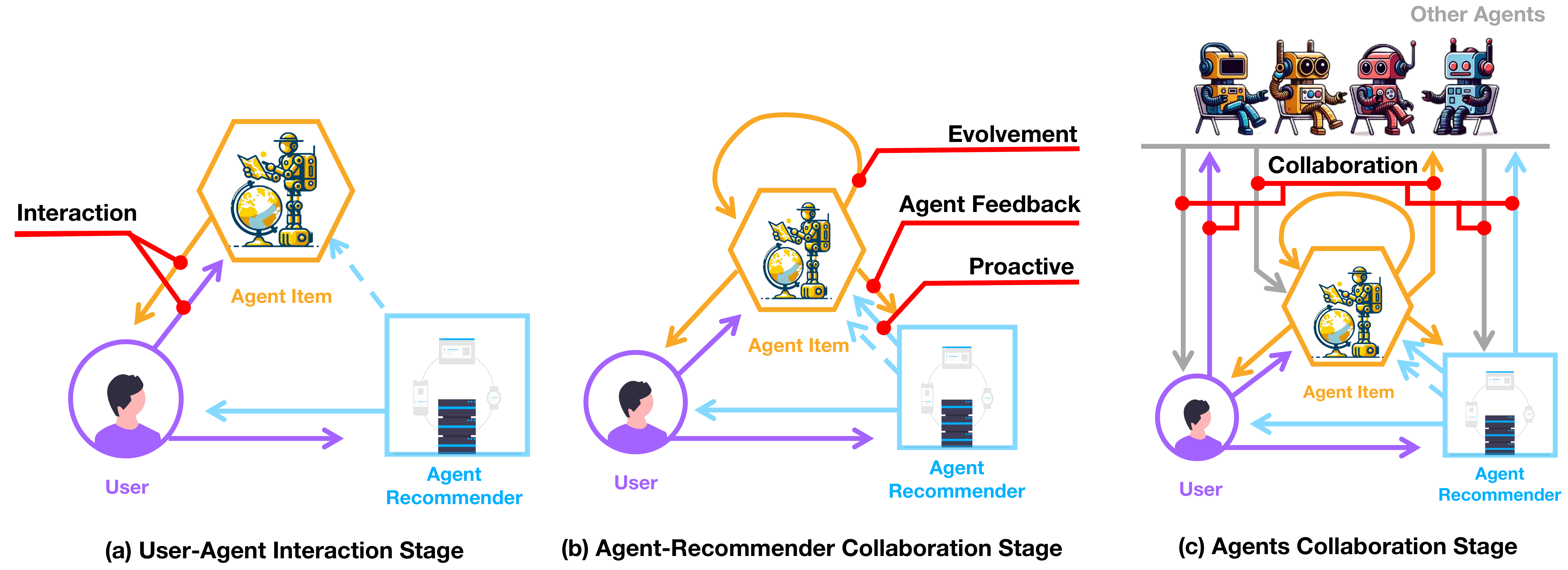}
\caption{The three stages of our proposed \name~paradigm.
(a) For the User-Agent interaction stage, 
users can interact efficiently with Agent items through natural language.
(b) For the Agent-Recommender collaboration stage,
Agent Item and Agent Recommender could interact with each other. 
``Evolvement'' means that the preference of the user can also be used for Agent Item to evolve itself or to evolve itself with the help of Agent Recommender.
``Agent Feedback'' refers to that the recommended Agent Item can feed the preference of the user back to Agent Recommender.
``Proactive'' stands for Agent Recommender can send information or issue instructions to Agent items.
(c) For the Agents collaboration stage, Agent Items can collaborate together to provide personalized information services for the user.
} 
\label{fig:three_stage_detail} 
\end{figure*}

\subsection{Potential Research Topics}
\label{sec:2.4}
Within \name, there exist numerous valuable research directions awaiting exploration. 
In this subsection, we point out several prospective and meaningful research topics:
\begin{itemize}[leftmargin=*]
    \item \textbf{Evaluation}.
    One crucial problem is how to evaluate the recommendation performance of \name~ since it significantly differs from existing recommender systems. 
    On the one hand, traditional recommendation datasets struggle to adapt to Rec4Agentverse, since Agent Item is quite different from previous items in the recommendation dataset. 
    On the other hand, existing evaluation metrics for recommendations also face challenges in applying to \name. 
    Existing recommendation metrics, such as NDCG and HR, are designed for traditional items in traditional recommender systems. 
    It is hard to accurately measure the user satisfaction for existing Agent Items with multi-round interactions. 
    Moreover, Agent Recommender may generate a new Agent Item for users or Agent Items may upgrade based on user feedback. 
    Evaluating user satisfaction with these new or upgraded Agent Items cannot purely rely on the user's implicit feedback such as interaction numbers. It needs to quantify the incremental performance compared to existing Agent Items. 
    
    \item \textbf{Preference Modeling}.
    How to effectively model users and enable \name~to provide users with a great personalized recommendation experience is also a pivotal problem. 
    On the one hand, it is crucial to explore effective methods for acquiring user preferences to provide users with desired recommendation results. 
    Several existing studies have indicated that modeling collaborative information poses challenges for LLM~\cite{bigrec, collm}. 
    Thus, it is worth investigating how to design efficient user modeling approaches for \name.
    On the other hand, there lies the challenge of effectively leveraging the collected user behavior data for training purposes. 
    The form of the user data with extensive natural language interactions and implicit feedback like dwell time and clicks gathered by \name~differs significantly from traditional recommendation data. 
    Moreover, the underlying architecture of the recommendation system in \name~is dissimilar to that of conventional models. 
    Consequently, devising a strategy to utilize this distinct data format in training the novel recommendation system within the \name~paradigm presents a formidable challenge.
    \item \textbf{Efficent Inference}.
    The \name~paradigm is based on LLM, which incurs significant inference costs~\cite{xia2023sheared, gu2023mamba}.
    The real-time inference requirements of recommender systems 
    give rise to research inquiries, such as how to mitigate the inferential costs of Rec4Agentverse without compromising its performance. 
    \item \textbf{Knowledge Update and Edit}.
    Due to the evolving distribution of knowledge in the real world~\cite{knowledge_survey_1, knowledge_survey_2}, \name~also faces the problem of updating its knowledge or editing out any incorrect information, posing a significant challenge. 
    These changes in knowledge distribution may arise from a shift in users' personal preferences, the new ~\itemname, or alterations in world knowledge. 
\end{itemize}

\subsection{Issues and Challenges}
\label{sec:2.5}
In this subsection, we shall delve into the potential issues and challenges of the \name~paradigm. 
\begin{itemize}[leftmargin=*]
    \item \textbf{Fairness and Bias}.
    \swt{The pre-training process of LLMs inherently involves incorporating data from the web, which may contain societal biases and unfair elements~\cite{fairness_1, fairness_2, jiang2024itemside}.
    \zjzz{Due to the social impact of the recommendation, fairness and bias are important issues in recommendation~\cite{fairness_rec_1, fairness_rec_2, fairness_rec_3}}.
    }
    \swt{Therefore, when employing the \name~paradigm, it becomes imperative to acknowledge and control the potential unfairness and bias in the recommended Agent Items and the information delivered by Agent Items, so as to mitigate the potential risks and negative societal impacts.} 
    \item \textbf{Privacy}.
    Users may inadvertently disclose their privacy while interacting with LLMs~\cite{li2023privacy, kim2023propile}. Since Rec4Agentverse is based on the LLM, safeguarding users' privacy will be an important challenge to address.
    On the one hand, the confidentiality of these user's private information must remain undisclosed to other users. 
    On the other hand, users should possess the utmost control over their own private data. 
    When a user requests that the model refrains from using their private data, \name~should proceed to unlearn such private information.
    \item \textbf{Harmfulness}.
    Agent Items may potentially generate harmful textual responses~\cite{align_1, align_2}, which deserve our attention in the application of \name. 
    Furthermore, Agent Items might be manipulated to execute harmful actions for users, for example, some fraudulent transactions. It is essential to regulate the harmfulness of \name in terms of generated content and executed actions. 
    \item \textbf{Robustness}.
    LLM may not be robust to the malicious prompt~\cite{yu2023gptfuzzer,lapid2023open}.
    In contrast to conventional recommendation systems, Rec4- Agentverse involves extensive use of prompts to interact with the user, requiring \name~to be robust to malicious prompts. \zjz{If \name~ is not sufficiently robust, it is susceptible to prompt injection attacks, or poisoning data that can lead to performance degradation or harmful output.}
    \item \textbf{Environmental Friendliness}.
    Considering the high energy consumption of training and inference LLMs~\cite{xia2023sheared, zhou2023opportunities}, ensuring the environmental friendliness of Rec4Agentverse is a crucial concern.
    We should try to save energy consumption and carbon emissions during the training, deployment, and inference processes in the \name.
\end{itemize}

\section{Discussion}
In this section, we contrast our proposed Rec4Agentverse paradigm with existing recommendation paradigms: retrieval-based recommendation and generative-based recommendation~\cite{wang2023generative}. Because of the distinctive characteristics of our paradigm's \itemname~and \recname, such as their powerful interactive capabilities and proactiveness in actively exploring user preferences, the Rec4Agentverse paradigm diverges from the traditional paradigms. Here, we delve into two transformative changes from the viewpoint of user preference modeling and collaborative mechanisms among the system's entities.


\paragraph{\textbf{User Preference Modeling:}} 
Beyond merely summarising user preference from passively received users' interactions on items like done in conventional paradigms, in our paradigm, both \recname~and \itemname~ could actively acquire information to enhance user preferences modeling. 
In traditional paradigms, the interactive capability of the recommender and items is limited, particularly for items such as movies and games that cannot engage in verbal communication. Consequently, user preference modeling for these paradigms typically relies on passively received feedback\footnote{One exception is the interactive recommendation, however, its item does not have the interactive capability.}. However, in our paradigm, both the recommender and item have the ability to actively interact with users through dialogue to directly acquire user preference information or collect further feedback for preference refinement, enhancing user preference modeling.


\paragraph{\textbf{Collaboration Mechanisms:}}

In our paradigm, collaboration between recommenders and items is becoming increasingly closer and more extensive. In contrast, traditional paradigms encounter challenges in actively fostering collaboration between items or between items and recommenders once an item is recommended. These enhanced collaborations undoubtedly elevate the service quality of both the \recname and \itemname. For instance, in our paradigm, when a recommended item falls short of fully meeting the user's needs due to its limitations, it can initiate communication with the recommender or collaborate with other Agent Items entities to address these shortcomings and better align with the user's preferences. Conversely, in traditional paradigms, users often need to turn to the recommender system for another recommendation, perpetuating the iterative process, which would diminish users' enthusiasm. Another example is that Agent Recommender can enrich the user profile by engaging in conversations with Agent Items that the user has interacted with in the past or is currently engaging with, thereby facilitating more effective recommendations.


Overall, these changes stem from the revolution of information flow dynamics in our paradigm, leading to a more decentralized approach to information management. 
In contrast to traditional paradigms where management is predominantly centralized within the recommender system, our paradigm grants greater autonomy to Agent Items in managing this process. 
This enables Agent Items to engage with users more independently and proactively, thereby improving their capacity to capture and manage user preference information effectively on the one hand, and enabling them to freely display information to users on the other. 
Simultaneously, this facilitates the exchange or feedback of such information among Agent Items or between Agent Items and Agent Recommender, fostering improved collaboration in final user modeling.

\begin{figure}[t]
  \centering
  \includegraphics[width=0.5\textwidth]{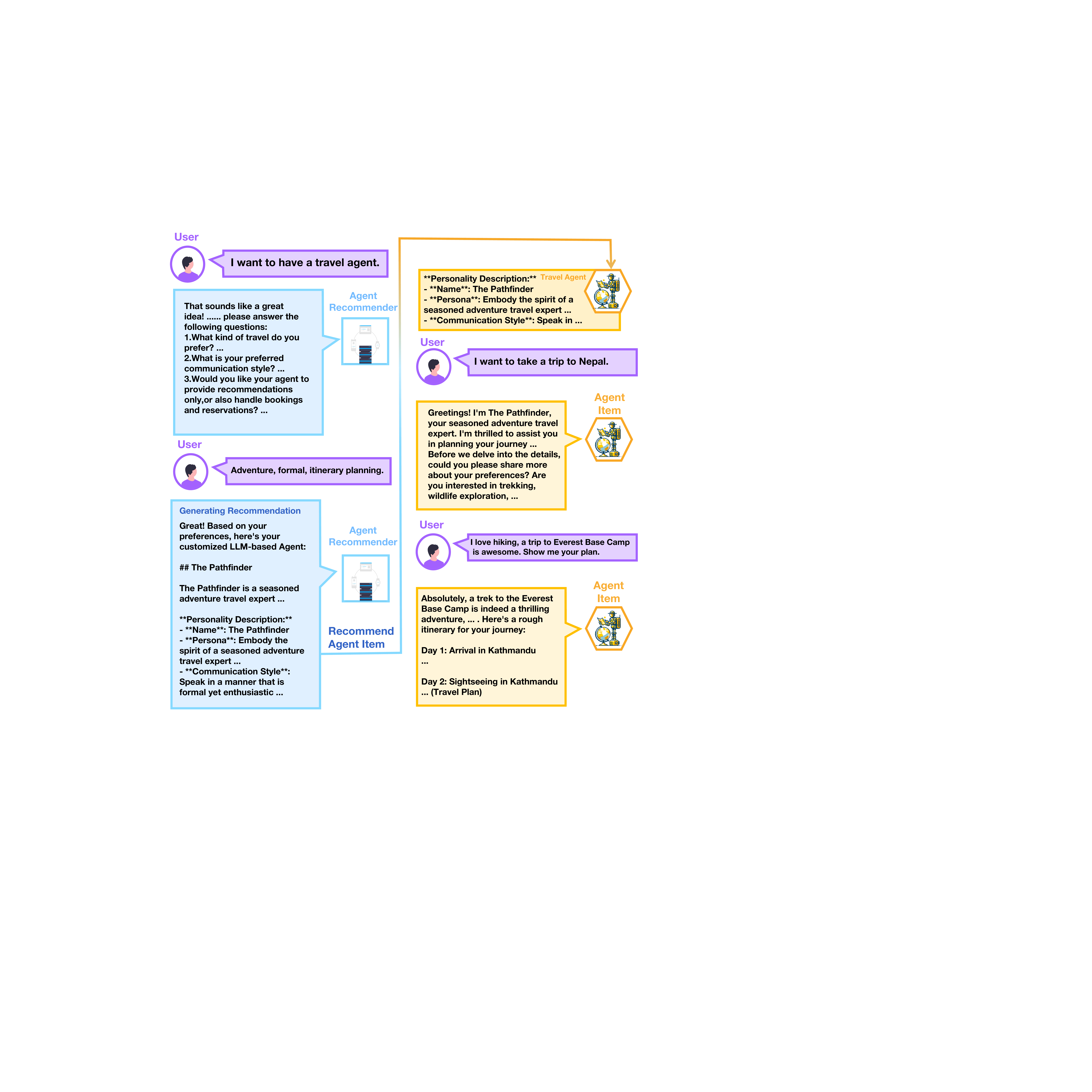}
  \caption{A case of the User-Agent interaction stage. 
The user expressed the desire for the Travel Agent to Agent Recommender and get back a recommendation. Subsequently, the user and the Travel Agent engaged in interactions to make the travel plan.}
  \label{fig:stage1_example}
\end{figure}

\begin{figure*}[t]
  \centering
  \includegraphics[width=\textwidth]{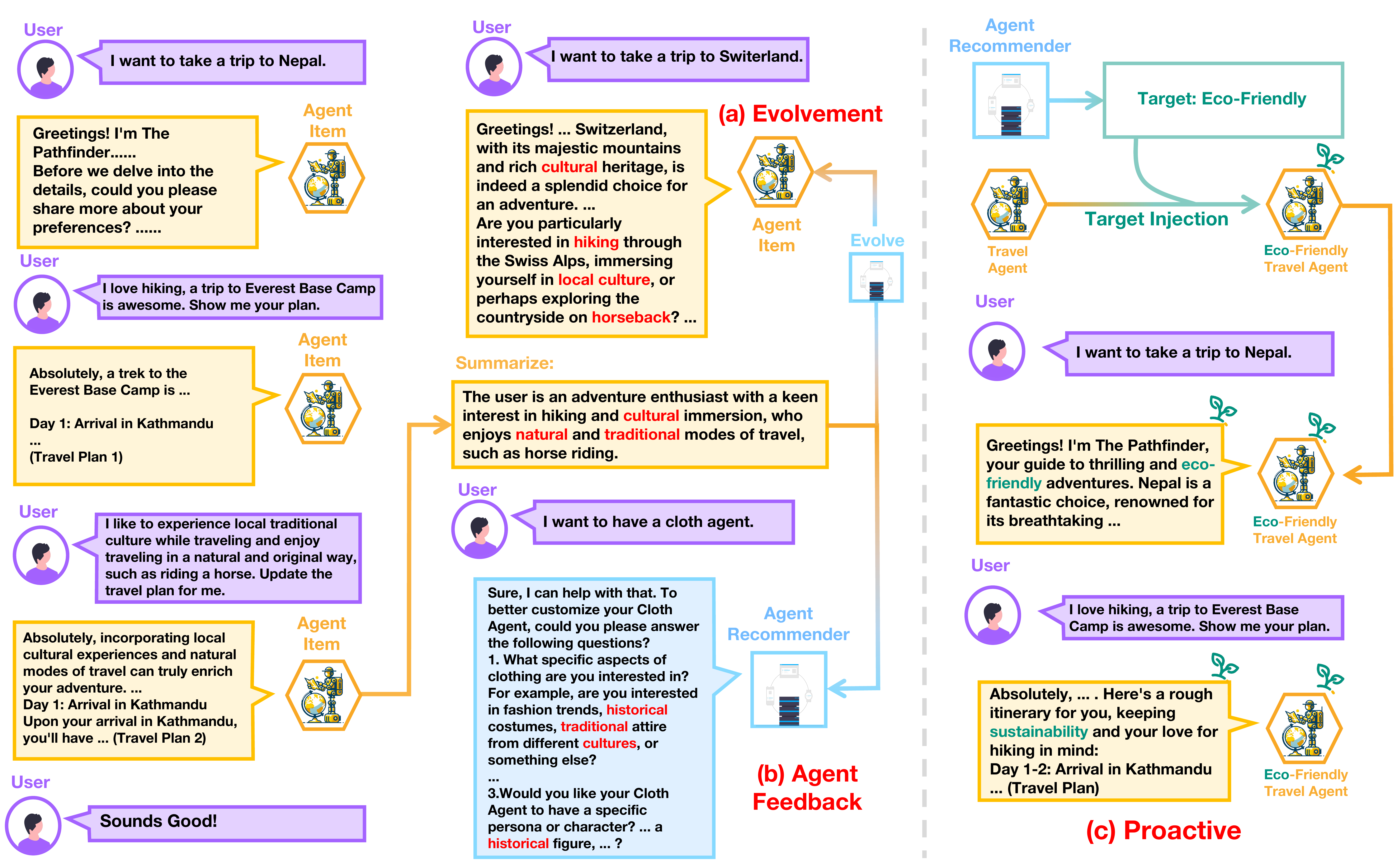}
  \caption{
Cases of three scenarios, namely Evolvement, Agent Feedback, and Proactive, at the Agent- Recommender Collaboration stage of Rec4Agentverse.
(a) For the Evolvement scenario, Agent Item has the ability to enhance itself with the help of Agent Recommender based on the user's preferences. (b) For the Agent Feedback scenario, 
Agent Item submits the user's preference to Agent Recommender so that Agent Recommender can provide better recommendations.
(c) For the Proactive scenario, Agent Recommender provides the eco-friendly target to Agent Item, and Agent Item successfully achieves the eco-friendly target in its interaction with the user.}
  \label{fig:stage2_1_example}
\end{figure*}
\section{Demonstration}

In this section, we explore the three stages of Rec4Agentverse through case studies, focusing on discussing the feasibility and potential formats of the paradigm. We present a case study involving a traveler who utilizes Rec4Agentverse throughout his/her journey, examining how Agent Recommender and Agent Item work and affect the user experience at each stage. This case study is based on ``gpt-4-32k''. Due to space constraints, we provide only the essential parts of the case study here, with additional details available at github\footnote{\zjzz{\url{https://github.com/jizhi-zhang/Rec4Agentverse_Case}}}. It's important to note that our case study serves as a preliminary indication of the feasibility of different stages within the Rec4Agentverse paradigm, and it does not fully encompass all the potential applications of our paradigm. 

\subsection{Stage 1: User-Agent Interaction}
In the User-Agent interaction stage, as shown in Figure~\ref{fig:three_stage_detail}(a), Agent Item primarily engages in interactions with the user, facilitating efficient information exchange between Agent Item and the user. To demonstrate this, we present a scenario where a user expresses their desire to travel to Nepal and interacts with an Agent Recommender and the recommended Travel Agent, as shown in Figure~\ref{fig:stage1_example}. 
The user initially seeks assistance from the Agent Recommender to find a Travel Agent. 
Upon inquiring about the user's preferences, the Agent Recommender customizes a Travel Agent specifically tailored to the user's needs. 
Subsequently, after further determining the user's interests, this Agent devises a comprehensive travel itinerary for the user. Therefore, there are main two information exchange flows: one between the user and Agent Recommender and one between the user and Agent item.

\subsubsection{\textbf{Information Flow between User and Agent Recommender}}
As depicted in Figure~\ref{fig:stage1_example}, in this example, in addition to passively receiving requests from the user, Agent Recommender could actively engage with the user to improve their recommendations. For instance, after the user expresses a desire to find a Travel Agent through dialogue, Agent Recommender could proactively pose questions to gain a more detailed high-level preference of the user about the travel. With additional feedback from the user, Agent Recommender could then provide accurate recommendations for a Travel Agent. This process bears some resemblance to traditional interactive recommendation methods.


\begin{figure*}[t]
  \centering
  \includegraphics[width=\textwidth]{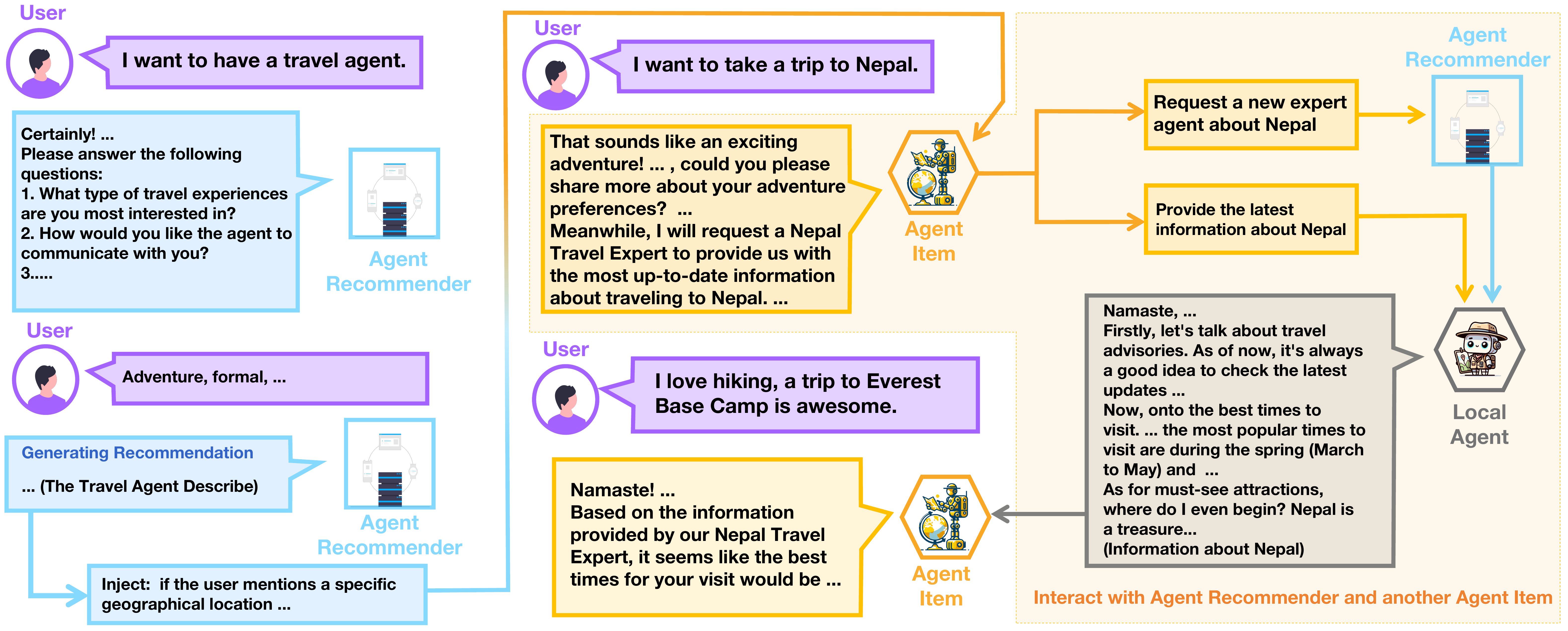}
  \caption{Preliminary case study of Agents Collaboration stage. 
  When the user asks about the travel plan for Nepal, the Travel Agent requires a specific Local Agent of Nepal from Agent Recommender to solve this problem. By conversation with the Local Agent about Nepal, the Travel Agent gets up-to-date information about Nepal which helps plan travel tours for the user.
  }
  \label{fig:stage3_example}
\end{figure*}

\subsubsection{\textbf{Information Flow between User and Agent Item}}

As illustrated in Figure~\ref{fig:stage1_example}, in stark contrast to the traditional paradigm, Agent Item is capable of interacting directly with the user. In our example, the Travel Agent initially learns about the user's interest in traveling to Nepal and their request for a travel plan. Subsequently, it could inquire further to uncover more specific preferences, obtaining the user's inclination to visit the ``Everest Base Camp''. This exchange of information allows Agent Item to develop a deeper understanding of the user's preferences, thereby enhancing its ability to provide tailored services to users.

\subsection{Stage 2: Agent-Recommender Collaboration}
%
In the agent-recommender collaboration stage, as depicted in Figure~\ref{fig:three_stage_detail}(b), there is potential for further information exchange between Agent Item and Agent Recommender. This exchange opens up three promising possibilities:
\begin{itemize}[leftmargin=*]
    \item \textit{\textbf{Evolvement}}: Agent Item can undergo evolution \zjzz{by itself or} with guidance from Agent Recommender
    .
    \item \textit{\textbf{Agent Feedback}}: Agent Item can provide valuable feedback to Agent Recommender.
    \item \textit{\textbf{Proactive}}: Agent Item can follow instructions from Agent Recommender to accomplish specific objectives.
\end{itemize} 
We illustrate these possibilities by extending the travel example, as depicted in Figure~\ref{fig:stage2_1_example}.


\subsubsection{\textbf{Evolvement}.}

Thanks to its ability to gather information from users and Agent Recommender, Agent Item can acquire valuable knowledge to achieve evolution, helping enhance future services. In the example illustrated in Figure~\ref{fig:stage2_1_example}, Agent Item can leverage the knowledge summarized by itself and obtained from Agent Recommender to achieve evolution. This evolution may involve improving its prompts, for instance. As a result, when the user makes their next request for a trip to a new destination, \eg Switzerland, the system will promptly present a travel itinerary that directly aligns with the user's personal preferences, taking into account their inclination towards ``\zjzz{hiking, cultural, and natural}'' experiences. This process of evolution enables the continuous tracking of user information and alleviates the burden on users to express their preferences in great detail in future interactions.


\subsubsection{\textbf{Agent Feedback}.}
Agent Item can also contribute feedback, namely agent feedback, to enhance the services of Agent Recommender in the future. In our example depicted in Figure~\ref{fig:stage2_1_example}, the recommended Travel Agent can provide a summarized understanding of the user's preferences, such as ``\zjzz{cultural, natural and so on}'', to Agent Recommender. Agent Recommender can then absorb this knowledge and improve its future services accordingly. Then, when a new request for a ``Cloth Agent'' arises, Agent Recommender can directly inquire whether the user is interested in environmentally friendly or culturally significant attire, based on the knowledge obtained from the Travel Agent. Through this information exchange, Agent Recommender can significantly enhance its services.



\subsubsection{\textbf{Proactive}}
Here, ``\zjzz{Proactive}'' refers to the ability of Agent Items to autonomously \zjzz{accomplish} specific objectives, which can originate from the Agent platform itself or aim to better align with user interests. 
\zjzz{An example is} shown in Figure~\ref{fig:stage2_1_example}, we assume that Agent Recommender has prior knowledge of the user's inclination towards eco-friendly options. 
Therefore, before the user initiates their interaction, Agent Recommender injects this eco-friendly objective into the recommended Travel Agent.
Consequently, when users engage with the Travel Agent, it will provide environmentally friendly travel options that fulfill the eco-friendly requirement. 
This proactive characteristic enhances user satisfaction and tailors the experience to their specific interests.

\subsection{Stage 3: Agents Collaboration}
 
Compared to the other two stages, the collaboration stage of the agents allows for further exchange of information among Agent Items, as depicted in Figure~\ref{fig:three_stage_detail}(c), enabling them to collaborate and enhance services for users. In the Travel Agent case illustrated in Figure~\ref{fig:stage3_example}, we present a potential example where multiple agents collaborate to complete the travel planning process.
Here's a step-by-step breakdown of the collaboration process:
\begin{itemize}[leftmargin=*]
    \item The user starts a conversation with Agent Recommender, expressing the desire to plan a travel tour.
    \item Agent Recommender suggests a Travel Agent whose goal is to help with travel tour planning.
    \item The user subsequently requests the Travel Agent to create a travel itinerary specifically tailored for Nepal.
    \item To acquire the latest information about Nepal, the Travel Agent sends a request to Agent Recommender for an Agent Item. This Agent Item should be able to provide up-to-date local information on Nepal, which will assist in creating the travel plan. 
    \item Agent Recommender responds by recommending a Local Agent who is knowledgeable about the current situation in Nepal.
    \item Ultimately, the Travel Agent integrates \zjzz{the current information about Nepal provided by the Local Agent into the travel itinerary design process to fulfill the user's needs. }
\end{itemize}
Our conclusion from the case suggests that by adopting a system of collaborative cooperation among agents, they are enabled to communicate more effectively and share information with each other. 
This exchange process significantly enriches their shared knowledge base. 
As a result, these agents are better equipped to address and cater to a more diverse and comprehensive range of user needs, thereby enhancing overall user satisfaction.

\section{Related Work}
\zjz{
In this section, we will mainly discuss two types of related work: LLM-based Recommendation and LLM-based Agent.
In the subsection on LLM-based Recommendation, we will emphasize the distinction between Rec4Agentverse and the current LLM-based Agent for recommendations.
}
\subsection{LLM for Recommendation}
With the emergence of powerful models like ChatGPT and their overwhelming advantages demonstrated in other domains~\cite{li2024towards, zhao2023survey}, an increasing number of researchers in the recommendation community have begun to explore the potential of applying LLMs to recommendation systems~\cite{llmrecsurvey1, llmrecsurvey2, llm_survey_3, lin2023multi, yang2023large}. 
This can be divided into two categories.
One category advocates for directly leveraging the capabilities of LLMs by utilizing appropriate prompts to stimulate their abilities in recommendation scenarios~\cite{junxu_rec, ChatgptGoodRec}. Some researchers directly employ LLMs for reordering~\cite{hou2023large}, while others distill the world knowledge of LLMs into traditional recommendation models to enhance recommendation performance~\cite{openworld_rec, wei2023llmrec}. Another group of researchers, however, believes that LLMs rarely encounter recommendation tasks during pre-training, and recommendation data often possess privatized characteristics~\cite{tallrec}. Therefore, there is an urgent need to explore the use of tuning methods to improve the recommendation performance of LLMs. In this regard, researchers have utilized instruction tuning to enable models to quickly learn various types of recommendation tasks in a generative manner~\cite{instructfollows, bigrec, llmconversationrec}, yielding significant improvements. Furthermore, it has been discovered that injecting collaborative information into LLMs through post-processing or modal injection methods can further enhance the performance of LLM-based Recommendations~\cite{collm, zheng2023adapting}.

\paragraph{\textbf{LLM-based Agents for Recommendation}}
Following the surge in popularity of LLMs, an increasing number of individuals exploring the use of LLMs to simulate social environments and perform various complex tasks~\cite{agent_survey_1, agent_survey_2}.
This has also promptly captured the attention of researchers in the field of recommender systems.
Specifically, some researchers aim to simulate users using agents, creating a virtual recommendation system environment (\eg RecAgent~\cite{recagent}, Agent4Rec~\cite{agent4rec}) to explore the social impact of recommendation algorithms within this simulated environment. 
Another group of researchers seeks to enhance recommendation performance through the use of agents. 
In this context, InteRecAgent~\cite{recommender_ai_agent} encapsulates LLMs as agents endowed with memory capabilities and scheduling proficiency, thereby enhancing recommendation performance.
AgentCF~\cite{agentcf} first attempts to optimize the self-introduction of both the User and Item by considering them as agents, and improving the self-introduction via User interaction with positive and negative Items.
The primary distinction between our work in this paper and the aforementioned research lies in their failure to transcend the confines of traditional recommendation paradigms.
Their objective remains confined to recommending passive items (\eg movies and games), indicating those items cannot actively interact with the user and obtain the user's intentions and preferences.
Conversely, as previously mentioned in this paper, we mainly talk about when we have employed an LLM-based Agent who is good at interaction, intelligent, and proactive for our recommended items.
Those agents can be viewed as a new type of information carrier and information dissemination tool and bring about changes in information systems.  

\subsection{LLM-based Agents}
LLM-based agents have been deployed in various domains to address a wide range of specific tasks, showcasing the robust and comprehensive capabilities of agents~\cite{agent_survey_1, agent_survey_2, chen2023agentverse}.
Some researchers are dedicated to exploring the upper limits of single-agent capabilities. They endeavor to utilize a solitary agent that can effectively adhere to a wide array of user instructions and successfully tackle a diverse range of complex tasks in both daily life~\cite{lifeagent1,lifeagent2,lifeagent3} and academic endeavors~\cite{academicagent1,academicagent2}.
Meanwhile, there is another faction of researchers who approach the birth of human social intelligence and believe that collective intelligence will bring about a more prosperous advancement~\cite{hao2023chatllm}. They endeavor to enhance the problem-solving capacity of such groups compared to individuals by means of collaborative efforts among multiple agents~\cite{mandi2023roco,wu2023autogen} or through agents engaging in mutual critique~\cite{adversal1, adversal2}.
In addition to these aspects, researchers are also devoted to exploring the interaction between the Agent and its environment to enhance the capabilities of the Agent. This encompasses the interaction between the Agent and humans to obtain human feedback~\cite{humanfeedback1,humanfeedback2,humanfeedback3} and the interaction between the Agent and the physical world through visual/audio modules to acquire additional knowledge~\cite{palme,Roboagent}, and so on.
In our Rec4Agentverse framework, as previously mentioned, we must delve into the profound application of recommendation systems.
This will enhance the ability of single agents to fulfill user needs, the collaborative capacity of multi-agents in aiding users, and the agents' ability to acquire feedback from the physical world and users themselves for self-evolution.

\section{Conclusion}

In this paper, we explore the application of recommender systems within an LLM-based Agents platform, examining how the unique characteristics of these agents - robust interactivity, intelligence, and proactiveness - alter the flow and presentation of information. This shift significantly impacts the function of recommendation systems, which are fundamentally designed to control information flow within systems.
We thus introduce a new paradigm, Rec4Agentverse, consisting of two elements: Agent Item and Agent Recommender. This paradigm is developed in three stages, each designed to enhance the interaction and information exchange among users, Agent Recommender, and Agent Items.
Then, we devised a simulated scenario to illustrate a user's need for travel planning assistance. 
Within this context, we conducted a comprehensive analysis of the three developmental phases of Rec4Agentverse. 
Our focus was on elucidating the unique attributes of each stage and exploring the scalability and potential of this paradigm. 
Through this simulation, we succinctly and clearly illustrate our understanding of the evolution of the system and its potential for future growth.

Moreover, it is important to note that Rec4Agentverse represents a novel recommendation paradigm that still requires extensive exploration in terms of its applicable fields, potential development directions, and existing risk issues. 
In light of this, we also delve into these aspects intending to inspire and encourage further advancements in the field.
Looking forward, we aim to investigate the Rec4Agentverse paradigm through quantitative research methods and explore strategies for its practical implementation.






\balance
\bibliographystyle{ACM-Reference-Format}
\bibliography{sample-base}


\begin{thebibliography}{65}


\ifx \showCODEN    \undefined \def \showCODEN     #1{\unskip}     \fi
\ifx \showDOI      \undefined \def \showDOI       #1{#1}\fi
\ifx \showISBNx    \undefined \def \showISBNx     #1{\unskip}     \fi
\ifx \showISBNxiii \undefined \def \showISBNxiii  #1{\unskip}     \fi
\ifx \showISSN     \undefined \def \showISSN      #1{\unskip}     \fi
\ifx \showLCCN     \undefined \def \showLCCN      #1{\unskip}     \fi
\ifx \shownote     \undefined \def \shownote      #1{#1}          \fi
\ifx \showarticletitle \undefined \def \showarticletitle #1{#1}   \fi
\ifx \showURL      \undefined \def \showURL       {\relax}        \fi
\providecommand\bibfield[2]{#2}
\providecommand\bibinfo[2]{#2}
\providecommand\natexlab[1]{#1}
\providecommand\showeprint[2][]{arXiv:#2}

\bibitem[Abbasian et~al\mbox{.}(2023)]%
        {diverse_agent_1}
\bibfield{author}{\bibinfo{person}{Mahyar Abbasian}, \bibinfo{person}{Iman Azimi}, \bibinfo{person}{Amir~M Rahmani}, {and} \bibinfo{person}{Ramesh Jain}.} \bibinfo{year}{2023}\natexlab{}.
\newblock \showarticletitle{Conversational health agents: A personalized llm-powered agent framework}.
\newblock \bibinfo{journal}{\emph{arXiv preprint arXiv:2310.02374}} (\bibinfo{year}{2023}).
\newblock


\bibitem[Bai et~al\mbox{.}(2022)]%
        {align_1}
\bibfield{author}{\bibinfo{person}{Yuntao Bai}, \bibinfo{person}{Saurav Kadavath}, \bibinfo{person}{Sandipan Kundu}, \bibinfo{person}{Amanda Askell}, \bibinfo{person}{Jackson Kernion}, \bibinfo{person}{Andy Jones}, \bibinfo{person}{Anna Chen}, \bibinfo{person}{Anna Goldie}, \bibinfo{person}{Azalia Mirhoseini}, \bibinfo{person}{Cameron McKinnon}, {et~al\mbox{.}}} \bibinfo{year}{2022}\natexlab{}.
\newblock \showarticletitle{Constitutional ai: Harmlessness from ai feedback}.
\newblock \bibinfo{journal}{\emph{arXiv preprint arXiv:2212.08073}} (\bibinfo{year}{2022}).
\newblock


\bibitem[Bakhtin et~al\mbox{.}(2022)]%
        {humanfeedback2}
\bibfield{author}{\bibinfo{person}{Anton Bakhtin}, \bibinfo{person}{David~J Wu}, \bibinfo{person}{Adam Lerer}, \bibinfo{person}{Jonathan Gray}, \bibinfo{person}{Athul~Paul Jacob}, \bibinfo{person}{Gabriele Farina}, \bibinfo{person}{Alexander~H Miller}, {and} \bibinfo{person}{Noam Brown}.} \bibinfo{year}{2022}\natexlab{}.
\newblock \showarticletitle{Mastering the game of no-press Diplomacy via human-regularized reinforcement learning and planning}.
\newblock \bibinfo{journal}{\emph{arXiv preprint arXiv:2210.05492}} (\bibinfo{year}{2022}).
\newblock


\bibitem[Bao et~al\mbox{.}(2023a)]%
        {bigrec}
\bibfield{author}{\bibinfo{person}{Keqin Bao}, \bibinfo{person}{Jizhi Zhang}, \bibinfo{person}{Wenjie Wang}, \bibinfo{person}{Yang Zhang}, \bibinfo{person}{Zhengyi Yang}, \bibinfo{person}{Yancheng Luo}, \bibinfo{person}{Fuli Feng}, \bibinfo{person}{Xiangnan He}, {and} \bibinfo{person}{Qi Tian}.} \bibinfo{year}{2023}\natexlab{a}.
\newblock \showarticletitle{A Bi-Step Grounding Paradigm for Large Language Models in Recommendation Systems}.
\newblock \bibinfo{journal}{\emph{arXiv preprint arXiv:2308.08434}} (\bibinfo{year}{2023}).
\newblock


\bibitem[Bao et~al\mbox{.}(2023b)]%
        {tallrec}
\bibfield{author}{\bibinfo{person}{Keqin Bao}, \bibinfo{person}{Jizhi Zhang}, \bibinfo{person}{Yang Zhang}, \bibinfo{person}{Wenjie Wang}, \bibinfo{person}{Fuli Feng}, {and} \bibinfo{person}{Xiangnan He}.} \bibinfo{year}{2023}\natexlab{b}.
\newblock \showarticletitle{TALLRec: An Effective and Efficient Tuning Framework to Align Large Language Model with Recommendation}. In \bibinfo{booktitle}{\emph{Proceedings of the 17th ACM Conference on Recommender Systems}} (Singapore, Singapore) \emph{(\bibinfo{series}{RecSys '23})}. \bibinfo{publisher}{Association for Computing Machinery}, \bibinfo{pages}{1007–1014}.
\newblock


\bibitem[Bharadhwaj et~al\mbox{.}(2023)]%
        {Roboagent}
\bibfield{author}{\bibinfo{person}{Homanga Bharadhwaj}, \bibinfo{person}{Jay Vakil}, \bibinfo{person}{Mohit Sharma}, \bibinfo{person}{Abhinav Gupta}, \bibinfo{person}{Shubham Tulsiani}, {and} \bibinfo{person}{Vikash Kumar}.} \bibinfo{year}{2023}\natexlab{}.
\newblock \showarticletitle{Roboagent: Generalization and efficiency in robot manipulation via semantic augmentations and action chunking}.
\newblock \bibinfo{journal}{\emph{arXiv preprint arXiv:2309.01918}} (\bibinfo{year}{2023}).
\newblock


\bibitem[Bran et~al\mbox{.}(2023)]%
        {academicagent1}
\bibfield{author}{\bibinfo{person}{Andres~M Bran}, \bibinfo{person}{Sam Cox}, \bibinfo{person}{Andrew~D White}, {and} \bibinfo{person}{Philippe Schwaller}.} \bibinfo{year}{2023}\natexlab{}.
\newblock \showarticletitle{ChemCrow: Augmenting large-language models with chemistry tools}.
\newblock \bibinfo{journal}{\emph{arXiv preprint arXiv:2304.05376}} (\bibinfo{year}{2023}).
\newblock


\bibitem[Chan et~al\mbox{.}(2023)]%
        {adversal1}
\bibfield{author}{\bibinfo{person}{Chi-Min Chan}, \bibinfo{person}{Weize Chen}, \bibinfo{person}{Yusheng Su}, \bibinfo{person}{Jianxuan Yu}, \bibinfo{person}{Wei Xue}, \bibinfo{person}{Shanghang Zhang}, \bibinfo{person}{Jie Fu}, {and} \bibinfo{person}{Zhiyuan Liu}.} \bibinfo{year}{2023}\natexlab{}.
\newblock \showarticletitle{Chateval: Towards better llm-based evaluators through multi-agent debate}.
\newblock \bibinfo{journal}{\emph{arXiv preprint arXiv:2308.07201}} (\bibinfo{year}{2023}).
\newblock


\bibitem[Chen et~al\mbox{.}(2023)]%
        {chen2023agentverse}
\bibfield{author}{\bibinfo{person}{Weize Chen}, \bibinfo{person}{Yusheng Su}, \bibinfo{person}{Jingwei Zuo}, \bibinfo{person}{Cheng Yang}, \bibinfo{person}{Chenfei Yuan}, \bibinfo{person}{Chen Qian}, \bibinfo{person}{Chi-Min Chan}, \bibinfo{person}{Yujia Qin}, \bibinfo{person}{Yaxi Lu}, \bibinfo{person}{Ruobing Xie}, {et~al\mbox{.}}} \bibinfo{year}{2023}\natexlab{}.
\newblock \showarticletitle{Agentverse: Facilitating multi-agent collaboration and exploring emergent behaviors in agents}.
\newblock \bibinfo{journal}{\emph{arXiv preprint arXiv:2308.10848}} (\bibinfo{year}{2023}).
\newblock


\bibitem[Dai et~al\mbox{.}(2023)]%
        {junxu_rec}
\bibfield{author}{\bibinfo{person}{Sunhao Dai} {et~al\mbox{.}}} \bibinfo{year}{2023}\natexlab{}.
\newblock \showarticletitle{Uncovering ChatGPT's Capabilities in Recommender Systems}. In \bibinfo{booktitle}{\emph{RecSys}}, \bibfield{editor}{\bibinfo{person}{Jie Zhang}, \bibinfo{person}{Li~Chen}, \bibinfo{person}{Shlomo Berkovsky}, \bibinfo{person}{Min Zhang}, \bibinfo{person}{Tommaso~Di Noia}, \bibinfo{person}{Justin Basilico}, \bibinfo{person}{Luiz Pizzato}, {and} \bibinfo{person}{Yang Song}} (Eds.). \bibinfo{publisher}{{ACM}}, \bibinfo{pages}{1126--1132}.
\newblock


\bibitem[Deldjoo(2024)]%
        {fairness_2}
\bibfield{author}{\bibinfo{person}{Yashar Deldjoo}.} \bibinfo{year}{2024}\natexlab{}.
\newblock \showarticletitle{Understanding Biases in ChatGPT-based Recommender Systems: Provider Fairness, Temporal Stability, and Recency}.
\newblock \bibinfo{journal}{\emph{arXiv preprint arXiv:2401.10545}} (\bibinfo{year}{2024}).
\newblock


\bibitem[Driess et~al\mbox{.}(2023)]%
        {palme}
\bibfield{author}{\bibinfo{person}{Danny Driess}, \bibinfo{person}{Fei Xia}, \bibinfo{person}{Mehdi~SM Sajjadi}, \bibinfo{person}{Corey Lynch}, \bibinfo{person}{Aakanksha Chowdhery}, \bibinfo{person}{Brian Ichter}, \bibinfo{person}{Ayzaan Wahid}, \bibinfo{person}{Jonathan Tompson}, \bibinfo{person}{Quan Vuong}, \bibinfo{person}{Tianhe Yu}, {et~al\mbox{.}}} \bibinfo{year}{2023}\natexlab{}.
\newblock \showarticletitle{Palm-e: An embodied multimodal language model}.
\newblock \bibinfo{journal}{\emph{arXiv preprint arXiv:2303.03378}} (\bibinfo{year}{2023}).
\newblock


\bibitem[(FAIR)† et~al\mbox{.}(2022)]%
        {humanfeedback1}
\bibfield{author}{\bibinfo{person}{Meta Fundamental AI Research Diplomacy~Team (FAIR)†}, \bibinfo{person}{Anton Bakhtin}, \bibinfo{person}{Noam Brown}, \bibinfo{person}{Emily Dinan}, \bibinfo{person}{Gabriele Farina}, \bibinfo{person}{Colin Flaherty}, \bibinfo{person}{Daniel Fried}, \bibinfo{person}{Andrew Goff}, \bibinfo{person}{Jonathan Gray}, \bibinfo{person}{Hengyuan Hu}, {et~al\mbox{.}}} \bibinfo{year}{2022}\natexlab{}.
\newblock \showarticletitle{Human-level play in the game of Diplomacy by combining language models with strategic reasoning}.
\newblock \bibinfo{journal}{\emph{Science}} \bibinfo{volume}{378}, \bibinfo{number}{6624} (\bibinfo{year}{2022}), \bibinfo{pages}{1067--1074}.
\newblock


\bibitem[Fan et~al\mbox{.}(2023)]%
        {llm_survey_3}
\bibfield{author}{\bibinfo{person}{Wenqi Fan}, \bibinfo{person}{Zihuai Zhao}, \bibinfo{person}{Jiatong Li}, \bibinfo{person}{Yunqing Liu}, \bibinfo{person}{Xiaowei Mei}, \bibinfo{person}{Yiqi Wang}, \bibinfo{person}{Jiliang Tang}, {and} \bibinfo{person}{Qing Li}.} \bibinfo{year}{2023}\natexlab{}.
\newblock \showarticletitle{Recommender systems in the era of large language models (llms)}.
\newblock \bibinfo{journal}{\emph{arXiv preprint arXiv:2307.02046}} (\bibinfo{year}{2023}).
\newblock


\bibitem[Feng et~al\mbox{.}(2023)]%
        {llmconversationrec}
\bibfield{author}{\bibinfo{person}{Yue Feng}, \bibinfo{person}{Shuchang Liu}, \bibinfo{person}{Zhenghai Xue}, \bibinfo{person}{Qingpeng Cai}, \bibinfo{person}{Lantao Hu}, \bibinfo{person}{Peng Jiang}, \bibinfo{person}{Kun Gai}, {and} \bibinfo{person}{Fei Sun}.} \bibinfo{year}{2023}\natexlab{}.
\newblock \showarticletitle{A Large Language Model Enhanced Conversational Recommender System}.
\newblock \bibinfo{journal}{\emph{CoRR}}  \bibinfo{volume}{abs/2308.06212} (\bibinfo{year}{2023}).
\newblock
\urldef\tempurl%
\url{https://doi.org/10.48550/arXiv.2308.06212}
\showDOI{\tempurl}
\showeprint[arXiv]{2308.06212}


\bibitem[Ge et~al\mbox{.}(2021)]%
        {fairness_rec_3}
\bibfield{author}{\bibinfo{person}{Yingqiang Ge}, \bibinfo{person}{Shuchang Liu}, \bibinfo{person}{Ruoyuan Gao}, \bibinfo{person}{Yikun Xian}, \bibinfo{person}{Yunqi Li}, \bibinfo{person}{Xiangyu Zhao}, \bibinfo{person}{Changhua Pei}, \bibinfo{person}{Fei Sun}, \bibinfo{person}{Junfeng Ge}, \bibinfo{person}{Wenwu Ou}, {and} \bibinfo{person}{Yongfeng Zhang}.} \bibinfo{year}{2021}\natexlab{}.
\newblock \showarticletitle{Towards long-term fairness in recommendation}. In \bibinfo{booktitle}{\emph{Proceedings of the 14th ACM international conference on web search and data mining}}. \bibinfo{pages}{445--453}.
\newblock


\bibitem[Gu and Dao(2023)]%
        {gu2023mamba}
\bibfield{author}{\bibinfo{person}{Albert Gu} {and} \bibinfo{person}{Tri Dao}.} \bibinfo{year}{2023}\natexlab{}.
\newblock \showarticletitle{Mamba: Linear-time sequence modeling with selective state spaces}.
\newblock \bibinfo{journal}{\emph{arXiv preprint arXiv:2312.00752}} (\bibinfo{year}{2023}).
\newblock


\bibitem[Gur et~al\mbox{.}(2023)]%
        {lifeagent1}
\bibfield{author}{\bibinfo{person}{Izzeddin Gur}, \bibinfo{person}{Hiroki Furuta}, \bibinfo{person}{Austin Huang}, \bibinfo{person}{Mustafa Safdari}, \bibinfo{person}{Yutaka Matsuo}, \bibinfo{person}{Douglas Eck}, {and} \bibinfo{person}{Aleksandra Faust}.} \bibinfo{year}{2023}\natexlab{}.
\newblock \showarticletitle{A real-world webagent with planning, long context understanding, and program synthesis}.
\newblock \bibinfo{journal}{\emph{arXiv preprint arXiv:2307.12856}} (\bibinfo{year}{2023}).
\newblock


\bibitem[Hao et~al\mbox{.}(2023)]%
        {hao2023chatllm}
\bibfield{author}{\bibinfo{person}{Rui Hao}, \bibinfo{person}{Linmei Hu}, \bibinfo{person}{Weijian Qi}, \bibinfo{person}{Qingliu Wu}, \bibinfo{person}{Yirui Zhang}, {and} \bibinfo{person}{Liqiang Nie}.} \bibinfo{year}{2023}\natexlab{}.
\newblock \showarticletitle{ChatLLM Network: More brains, More intelligence}.
\newblock \bibinfo{journal}{\emph{arXiv preprint arXiv:2304.12998}} (\bibinfo{year}{2023}).
\newblock


\bibitem[Hou et~al\mbox{.}(2023)]%
        {hou2023large}
\bibfield{author}{\bibinfo{person}{Yupeng Hou}, \bibinfo{person}{Junjie Zhang}, \bibinfo{person}{Zihan Lin}, \bibinfo{person}{Hongyu Lu}, \bibinfo{person}{Ruobing Xie}, \bibinfo{person}{Julian McAuley}, {and} \bibinfo{person}{Wayne~Xin Zhao}.} \bibinfo{year}{2023}\natexlab{}.
\newblock \showarticletitle{Large language models are zero-shot rankers for recommender systems}.
\newblock \bibinfo{journal}{\emph{arXiv preprint arXiv:2305.08845}} (\bibinfo{year}{2023}).
\newblock


\bibitem[Huang et~al\mbox{.}(2023)]%
        {recommender_ai_agent}
\bibfield{author}{\bibinfo{person}{Xu Huang}, \bibinfo{person}{Jianxun Lian}, \bibinfo{person}{Yuxuan Lei}, \bibinfo{person}{Jing Yao}, \bibinfo{person}{Defu Lian}, {and} \bibinfo{person}{Xing Xie}.} \bibinfo{year}{2023}\natexlab{}.
\newblock \showarticletitle{Recommender ai agent: Integrating large language models for interactive recommendations}.
\newblock \bibinfo{journal}{\emph{arXiv preprint arXiv:2308.16505}} (\bibinfo{year}{2023}).
\newblock


\bibitem[Ji et~al\mbox{.}(2023)]%
        {align_2}
\bibfield{author}{\bibinfo{person}{Jiaming Ji}, \bibinfo{person}{Tianyi Qiu}, \bibinfo{person}{Boyuan Chen}, \bibinfo{person}{Borong Zhang}, \bibinfo{person}{Hantao Lou}, \bibinfo{person}{Kaile Wang}, \bibinfo{person}{Yawen Duan}, \bibinfo{person}{Zhonghao He}, \bibinfo{person}{Jiayi Zhou}, \bibinfo{person}{Zhaowei Zhang}, {et~al\mbox{.}}} \bibinfo{year}{2023}\natexlab{}.
\newblock \showarticletitle{Ai alignment: A comprehensive survey}.
\newblock \bibinfo{journal}{\emph{arXiv preprint arXiv:2310.19852}} (\bibinfo{year}{2023}).
\newblock


\bibitem[Jiang et~al\mbox{.}(2024)]%
        {jiang2024itemside}
\bibfield{author}{\bibinfo{person}{Meng Jiang}, \bibinfo{person}{Keqin Bao}, \bibinfo{person}{Jizhi Zhang}, \bibinfo{person}{Wenjie Wang}, \bibinfo{person}{Zhengyi Yang}, \bibinfo{person}{Fuli Feng}, {and} \bibinfo{person}{Xiangnan He}.} \bibinfo{year}{2024}\natexlab{}.
\newblock \showarticletitle{Item-side Fairness of Large Language Model-based Recommendation System}.
\newblock \bibinfo{journal}{\emph{arXiv preprint arXiv:2402.15215}} (\bibinfo{year}{2024}).
\newblock


\bibitem[Kang and Kim(2023)]%
        {academicagent2}
\bibfield{author}{\bibinfo{person}{Yeonghun Kang} {and} \bibinfo{person}{Jihan Kim}.} \bibinfo{year}{2023}\natexlab{}.
\newblock \showarticletitle{Chatmof: An autonomous ai system for predicting and generating metal-organic frameworks}.
\newblock \bibinfo{journal}{\emph{arXiv preprint arXiv:2308.01423}} (\bibinfo{year}{2023}).
\newblock


\bibitem[Kim et~al\mbox{.}(2023)]%
        {kim2023propile}
\bibfield{author}{\bibinfo{person}{Siwon Kim}, \bibinfo{person}{Sangdoo Yun}, \bibinfo{person}{Hwaran Lee}, \bibinfo{person}{Martin Gubri}, \bibinfo{person}{Sungroh Yoon}, {and} \bibinfo{person}{Seong~Joon Oh}.} \bibinfo{year}{2023}\natexlab{}.
\newblock \showarticletitle{Propile: Probing privacy leakage in large language models}.
\newblock \bibinfo{journal}{\emph{arXiv preprint arXiv:2307.01881}} (\bibinfo{year}{2023}).
\newblock


\bibitem[Lapid et~al\mbox{.}(2023)]%
        {lapid2023open}
\bibfield{author}{\bibinfo{person}{Raz Lapid}, \bibinfo{person}{Ron Langberg}, {and} \bibinfo{person}{Moshe Sipper}.} \bibinfo{year}{2023}\natexlab{}.
\newblock \showarticletitle{Open sesame! universal black box jailbreaking of large language models}.
\newblock \bibinfo{journal}{\emph{arXiv preprint arXiv:2309.01446}} (\bibinfo{year}{2023}).
\newblock


\bibitem[Li et~al\mbox{.}(2023c)]%
        {humanfeedback3}
\bibfield{author}{\bibinfo{person}{Guohao Li}, \bibinfo{person}{Hasan Abed Al~Kader Hammoud}, \bibinfo{person}{Hani Itani}, \bibinfo{person}{Dmitrii Khizbullin}, {and} \bibinfo{person}{Bernard Ghanem}.} \bibinfo{year}{2023}\natexlab{c}.
\newblock \showarticletitle{Camel: Communicative agents for" mind" exploration of large scale language model society}.
\newblock \bibinfo{journal}{\emph{arXiv preprint arXiv:2303.17760}} (\bibinfo{year}{2023}).
\newblock


\bibitem[Li et~al\mbox{.}(2023a)]%
        {li2023privacy}
\bibfield{author}{\bibinfo{person}{Haoran Li}, \bibinfo{person}{Yulin Chen}, \bibinfo{person}{Jinglong Luo}, \bibinfo{person}{Yan Kang}, \bibinfo{person}{Xiaojin Zhang}, \bibinfo{person}{Qi Hu}, \bibinfo{person}{Chunkit Chan}, {and} \bibinfo{person}{Yangqiu Song}.} \bibinfo{year}{2023}\natexlab{a}.
\newblock \showarticletitle{Privacy in large language models: Attacks, defenses and future directions}.
\newblock \bibinfo{journal}{\emph{arXiv preprint arXiv:2310.10383}} (\bibinfo{year}{2023}).
\newblock


\bibitem[Li et~al\mbox{.}(2024)]%
        {li2024towards}
\bibfield{author}{\bibinfo{person}{Sihang Li}, \bibinfo{person}{Zhiyuan Liu}, \bibinfo{person}{Yanchen Luo}, \bibinfo{person}{Xiang Wang}, \bibinfo{person}{Xiangnan He}, \bibinfo{person}{Kenji Kawaguchi}, \bibinfo{person}{Tat-Seng Chua}, {and} \bibinfo{person}{Qi Tian}.} \bibinfo{year}{2024}\natexlab{}.
\newblock \showarticletitle{Towards 3D Molecule-Text Interpretation in Language Models}.
\newblock \bibinfo{journal}{\emph{ICLR}} (\bibinfo{year}{2024}).
\newblock


\bibitem[Li et~al\mbox{.}(2023b)]%
        {fairness_rec_1}
\bibfield{author}{\bibinfo{person}{Yunqi Li}, \bibinfo{person}{Hanxiong Chen}, \bibinfo{person}{Shuyuan Xu}, \bibinfo{person}{Yingqiang Ge}, \bibinfo{person}{Juntao Tan}, \bibinfo{person}{Shuchang Liu}, {and} \bibinfo{person}{Yongfeng Zhang}.} \bibinfo{year}{2023}\natexlab{b}.
\newblock \showarticletitle{Fairness in recommendation: Foundations, methods, and applications}.
\newblock \bibinfo{journal}{\emph{ACM Transactions on Intelligent Systems and Technology}} (\bibinfo{year}{2023}).
\newblock


\bibitem[Lin et~al\mbox{.}(2023a)]%
        {llmrecsurvey2}
\bibfield{author}{\bibinfo{person}{Jianghao Lin}, \bibinfo{person}{Xinyi Dai}, \bibinfo{person}{Yunjia Xi}, \bibinfo{person}{Weiwen Liu}, \bibinfo{person}{Bo Chen}, \bibinfo{person}{Xiangyang Li}, \bibinfo{person}{Chenxu Zhu}, \bibinfo{person}{Huifeng Guo}, \bibinfo{person}{Yong Yu}, \bibinfo{person}{Ruiming Tang}, {et~al\mbox{.}}} \bibinfo{year}{2023}\natexlab{a}.
\newblock \showarticletitle{How Can Recommender Systems Benefit from Large Language Models: A Survey}.
\newblock \bibinfo{journal}{\emph{arXiv preprint arXiv:2306.05817}} (\bibinfo{year}{2023}).
\newblock


\bibitem[Lin et~al\mbox{.}(2023b)]%
        {lin2023multi}
\bibfield{author}{\bibinfo{person}{Xinyu Lin}, \bibinfo{person}{Wenjie Wang}, \bibinfo{person}{Yongqi Li}, \bibinfo{person}{Fuli Feng}, \bibinfo{person}{See-Kiong Ng}, {and} \bibinfo{person}{Tat-Seng Chua}.} \bibinfo{year}{2023}\natexlab{b}.
\newblock \showarticletitle{A multi-facet paradigm to bridge large language model and recommendation}.
\newblock \bibinfo{journal}{\emph{arXiv preprint arXiv:2310.06491}} (\bibinfo{year}{2023}).
\newblock


\bibitem[Lin and Chen(2023)]%
        {lin2023llm}
\bibfield{author}{\bibinfo{person}{Yen-Ting Lin} {and} \bibinfo{person}{Yun-Nung Chen}.} \bibinfo{year}{2023}\natexlab{}.
\newblock \showarticletitle{LLM-Eval: Unified Multi-Dimensional Automatic Evaluation for Open-Domain Conversations with Large Language Models}.
\newblock \bibinfo{journal}{\emph{arXiv preprint arXiv:2305.13711}} (\bibinfo{year}{2023}).
\newblock


\bibitem[Liu et~al\mbox{.}(2023a)]%
        {plan_1}
\bibfield{author}{\bibinfo{person}{Bo Liu}, \bibinfo{person}{Yuqian Jiang}, \bibinfo{person}{Xiaohan Zhang}, \bibinfo{person}{Qiang Liu}, \bibinfo{person}{Shiqi Zhang}, \bibinfo{person}{Joydeep Biswas}, {and} \bibinfo{person}{Peter Stone}.} \bibinfo{year}{2023}\natexlab{a}.
\newblock \showarticletitle{Llm+ p: Empowering large language models with optimal planning proficiency}.
\newblock \bibinfo{journal}{\emph{arXiv preprint arXiv:2304.11477}} (\bibinfo{year}{2023}).
\newblock


\bibitem[Liu et~al\mbox{.}(2023b)]%
        {ChatgptGoodRec}
\bibfield{author}{\bibinfo{person}{Junling Liu}, \bibinfo{person}{Chao Liu}, \bibinfo{person}{Renjie Lv}, \bibinfo{person}{Kang Zhou}, {and} \bibinfo{person}{Yan Zhang}.} \bibinfo{year}{2023}\natexlab{b}.
\newblock \showarticletitle{Is chatgpt a good recommender? a preliminary study}.
\newblock \bibinfo{journal}{\emph{arXiv preprint arXiv:2304.10149}} (\bibinfo{year}{2023}).
\newblock


\bibitem[Mandi et~al\mbox{.}(2023)]%
        {mandi2023roco}
\bibfield{author}{\bibinfo{person}{Zhao Mandi}, \bibinfo{person}{Shreeya Jain}, {and} \bibinfo{person}{Shuran Song}.} \bibinfo{year}{2023}\natexlab{}.
\newblock \showarticletitle{Roco: Dialectic multi-robot collaboration with large language models}.
\newblock \bibinfo{journal}{\emph{arXiv preprint arXiv:2307.04738}} (\bibinfo{year}{2023}).
\newblock


\bibitem[Park et~al\mbox{.}(2023)]%
        {park2023generative}
\bibfield{author}{\bibinfo{person}{Joon~Sung Park}, \bibinfo{person}{Joseph O'Brien}, \bibinfo{person}{Carrie~Jun Cai}, \bibinfo{person}{Meredith~Ringel Morris}, \bibinfo{person}{Percy Liang}, {and} \bibinfo{person}{Michael~S Bernstein}.} \bibinfo{year}{2023}\natexlab{}.
\newblock \showarticletitle{Generative agents: Interactive simulacra of human behavior}. In \bibinfo{booktitle}{\emph{Proceedings of the 36th Annual ACM Symposium on User Interface Software and Technology}}. \bibinfo{pages}{1--22}.
\newblock


\bibitem[Qian et~al\mbox{.}(2023)]%
        {task_1}
\bibfield{author}{\bibinfo{person}{Chen Qian}, \bibinfo{person}{Xin Cong}, \bibinfo{person}{Cheng Yang}, \bibinfo{person}{Weize Chen}, \bibinfo{person}{Yusheng Su}, \bibinfo{person}{Juyuan Xu}, \bibinfo{person}{Zhiyuan Liu}, {and} \bibinfo{person}{Maosong Sun}.} \bibinfo{year}{2023}\natexlab{}.
\newblock \showarticletitle{Communicative agents for software development}.
\newblock \bibinfo{journal}{\emph{arXiv preprint arXiv:2307.07924}} (\bibinfo{year}{2023}).
\newblock


\bibitem[Shao et~al\mbox{.}(2023)]%
        {shao2023character}
\bibfield{author}{\bibinfo{person}{Yunfan Shao}, \bibinfo{person}{Linyang Li}, \bibinfo{person}{Junqi Dai}, {and} \bibinfo{person}{Xipeng Qiu}.} \bibinfo{year}{2023}\natexlab{}.
\newblock \showarticletitle{Character-llm: A trainable agent for role-playing}.
\newblock \bibinfo{journal}{\emph{arXiv preprint arXiv:2310.10158}} (\bibinfo{year}{2023}).
\newblock


\bibitem[Song et~al\mbox{.}(2023)]%
        {song2023llm}
\bibfield{author}{\bibinfo{person}{Chan~Hee Song}, \bibinfo{person}{Jiaman Wu}, \bibinfo{person}{Clayton Washington}, \bibinfo{person}{Brian~M Sadler}, \bibinfo{person}{Wei-Lun Chao}, {and} \bibinfo{person}{Yu Su}.} \bibinfo{year}{2023}\natexlab{}.
\newblock \showarticletitle{Llm-planner: Few-shot grounded planning for embodied agents with large language models}. In \bibinfo{booktitle}{\emph{Proceedings of the IEEE/CVF International Conference on Computer Vision}}. \bibinfo{pages}{2998--3009}.
\newblock


\bibitem[Wang et~al\mbox{.}(2023d)]%
        {lifeagent3}
\bibfield{author}{\bibinfo{person}{Guanzhi Wang}, \bibinfo{person}{Yuqi Xie}, \bibinfo{person}{Yunfan Jiang}, \bibinfo{person}{Ajay Mandlekar}, \bibinfo{person}{Chaowei Xiao}, \bibinfo{person}{Yuke Zhu}, \bibinfo{person}{Linxi Fan}, {and} \bibinfo{person}{Anima Anandkumar}.} \bibinfo{year}{2023}\natexlab{d}.
\newblock \showarticletitle{Voyager: An open-ended embodied agent with large language models}.
\newblock \bibinfo{journal}{\emph{arXiv preprint arXiv:2305.16291}} (\bibinfo{year}{2023}).
\newblock


\bibitem[Wang et~al\mbox{.}(2023b)]%
        {agent_survey_1}
\bibfield{author}{\bibinfo{person}{Lei Wang}, \bibinfo{person}{Chen Ma}, \bibinfo{person}{Xueyang Feng}, \bibinfo{person}{Zeyu Zhang}, \bibinfo{person}{Hao Yang}, \bibinfo{person}{Jingsen Zhang}, \bibinfo{person}{Zhiyuan Chen}, \bibinfo{person}{Jiakai Tang}, \bibinfo{person}{Xu Chen}, \bibinfo{person}{Yankai Lin}, \bibinfo{person}{Wayne~Xin Zhao}, \bibinfo{person}{Zhewei Wei}, {and} \bibinfo{person}{Ji{-}Rong Wen}.} \bibinfo{year}{2023}\natexlab{b}.
\newblock \showarticletitle{A Survey on Large Language Model based Autonomous Agents}.
\newblock \bibinfo{journal}{\emph{CoRR}}  \bibinfo{volume}{abs/2308.11432} (\bibinfo{year}{2023}).
\newblock
\urldef\tempurl%
\url{https://doi.org/10.48550/ARXIV.2308.11432}
\showDOI{\tempurl}
\showeprint[arXiv]{2308.11432}


\bibitem[Wang et~al\mbox{.}(2023e)]%
        {recagent}
\bibfield{author}{\bibinfo{person}{Lei Wang}, \bibinfo{person}{Jingsen Zhang}, \bibinfo{person}{Xu Chen}, \bibinfo{person}{Yankai Lin}, \bibinfo{person}{Ruihua Song}, \bibinfo{person}{Wayne~Xin Zhao}, {and} \bibinfo{person}{Ji-Rong Wen}.} \bibinfo{year}{2023}\natexlab{e}.
\newblock \showarticletitle{RecAgent: A Novel Simulation Paradigm for Recommender Systems}.
\newblock \bibinfo{journal}{\emph{arXiv preprint arXiv:2306.02552}} (\bibinfo{year}{2023}).
\newblock


\bibitem[Wang et~al\mbox{.}(2023f)]%
        {knowledge_survey_1}
\bibfield{author}{\bibinfo{person}{Song Wang}, \bibinfo{person}{Yaochen Zhu}, \bibinfo{person}{Haochen Liu}, \bibinfo{person}{Zaiyi Zheng}, \bibinfo{person}{Chen Chen}, {et~al\mbox{.}}} \bibinfo{year}{2023}\natexlab{f}.
\newblock \showarticletitle{Knowledge editing for large language models: A survey}.
\newblock \bibinfo{journal}{\emph{arXiv preprint arXiv:2310.16218}} (\bibinfo{year}{2023}).
\newblock


\bibitem[Wang et~al\mbox{.}(2023a)]%
        {wang2023generative}
\bibfield{author}{\bibinfo{person}{Wenjie Wang}, \bibinfo{person}{Xinyu Lin}, \bibinfo{person}{Fuli Feng}, \bibinfo{person}{Xiangnan He}, {and} \bibinfo{person}{Tat-Seng Chua}.} \bibinfo{year}{2023}\natexlab{a}.
\newblock \showarticletitle{Generative recommendation: Towards next-generation recommender paradigm}.
\newblock \bibinfo{journal}{\emph{arXiv preprint arXiv:2304.03516}} (\bibinfo{year}{2023}).
\newblock


\bibitem[Wang et~al\mbox{.}(2023c)]%
        {fairness_rec_2}
\bibfield{author}{\bibinfo{person}{Yifan Wang}, \bibinfo{person}{Weizhi Ma}, \bibinfo{person}{Min Zhang}, \bibinfo{person}{Yiqun Liu}, {and} \bibinfo{person}{Shaoping Ma}.} \bibinfo{year}{2023}\natexlab{c}.
\newblock \showarticletitle{A survey on the fairness of recommender systems}.
\newblock \bibinfo{journal}{\emph{ACM Transactions on Information Systems}} (\bibinfo{year}{2023}).
\newblock


\bibitem[Wei et~al\mbox{.}(2023)]%
        {wei2023llmrec}
\bibfield{author}{\bibinfo{person}{Wei Wei}, \bibinfo{person}{Xubin Ren}, \bibinfo{person}{Jiabin Tang}, \bibinfo{person}{Qinyong Wang}, \bibinfo{person}{Lixin Su}, \bibinfo{person}{Suqi Cheng}, \bibinfo{person}{Junfeng Wang}, \bibinfo{person}{Dawei Yin}, {and} \bibinfo{person}{Chao Huang}.} \bibinfo{year}{2023}\natexlab{}.
\newblock \showarticletitle{Llmrec: Large language models with graph augmentation for recommendation}.
\newblock \bibinfo{journal}{\emph{arXiv preprint arXiv:2311.00423}} (\bibinfo{year}{2023}).
\newblock


\bibitem[Wu et~al\mbox{.}(2023b)]%
        {llmrecsurvey1}
\bibfield{author}{\bibinfo{person}{Likang Wu}, \bibinfo{person}{Zhi Zheng}, \bibinfo{person}{Zhaopeng Qiu}, \bibinfo{person}{Hao Wang}, \bibinfo{person}{Hongchao Gu}, \bibinfo{person}{Tingjia Shen}, \bibinfo{person}{Chuan Qin}, \bibinfo{person}{Chen Zhu}, \bibinfo{person}{Hengshu Zhu}, \bibinfo{person}{Qi Liu}, {et~al\mbox{.}}} \bibinfo{year}{2023}\natexlab{b}.
\newblock \showarticletitle{A Survey on Large Language Models for Recommendation}.
\newblock \bibinfo{journal}{\emph{arXiv preprint arXiv:2305.19860}} (\bibinfo{year}{2023}).
\newblock


\bibitem[Wu et~al\mbox{.}(2023a)]%
        {wu2023autogen}
\bibfield{author}{\bibinfo{person}{Qingyun Wu}, \bibinfo{person}{Gagan Bansal}, \bibinfo{person}{Jieyu Zhang}, \bibinfo{person}{Yiran Wu}, \bibinfo{person}{Shaokun Zhang}, \bibinfo{person}{Erkang Zhu}, \bibinfo{person}{Beibin Li}, \bibinfo{person}{Li Jiang}, \bibinfo{person}{Xiaoyun Zhang}, {and} \bibinfo{person}{Chi Wang}.} \bibinfo{year}{2023}\natexlab{a}.
\newblock \showarticletitle{Autogen: Enabling next-gen llm applications via multi-agent conversation framework}.
\newblock \bibinfo{journal}{\emph{arXiv preprint arXiv:2308.08155}} (\bibinfo{year}{2023}).
\newblock


\bibitem[Xi et~al\mbox{.}(2023b)]%
        {openworld_rec}
\bibfield{author}{\bibinfo{person}{Yunjia Xi}, \bibinfo{person}{Weiwen Liu}, \bibinfo{person}{Jianghao Lin}, \bibinfo{person}{Jieming Zhu}, \bibinfo{person}{Bo Chen}, \bibinfo{person}{Ruiming Tang}, \bibinfo{person}{Weinan Zhang}, \bibinfo{person}{Rui Zhang}, {and} \bibinfo{person}{Yong Yu}.} \bibinfo{year}{2023}\natexlab{b}.
\newblock \showarticletitle{Towards Open-World Recommendation with Knowledge Augmentation from Large Language Models}.
\newblock \bibinfo{journal}{\emph{CoRR}}  \bibinfo{volume}{abs/2306.10933} (\bibinfo{year}{2023}).
\newblock


\bibitem[Xi et~al\mbox{.}(2023a)]%
        {agent_survey_2}
\bibfield{author}{\bibinfo{person}{Zhiheng Xi}, \bibinfo{person}{Wenxiang Chen}, \bibinfo{person}{Xin Guo}, \bibinfo{person}{Wei He}, \bibinfo{person}{Yiwen Ding}, \bibinfo{person}{Boyang Hong}, \bibinfo{person}{Ming Zhang}, \bibinfo{person}{Junzhe Wang}, \bibinfo{person}{Senjie Jin}, \bibinfo{person}{Enyu Zhou}, \bibinfo{person}{Rui Zheng}, \bibinfo{person}{Xiaoran Fan}, \bibinfo{person}{Xiao Wang}, \bibinfo{person}{Limao Xiong}, \bibinfo{person}{Yuhao Zhou}, \bibinfo{person}{Weiran Wang}, \bibinfo{person}{Changhao Jiang}, \bibinfo{person}{Yicheng Zou}, \bibinfo{person}{Xiangyang Liu}, \bibinfo{person}{Zhangyue Yin}, \bibinfo{person}{Shihan Dou}, \bibinfo{person}{Rongxiang Weng}, \bibinfo{person}{Wensen Cheng}, \bibinfo{person}{Qi Zhang}, \bibinfo{person}{Wenjuan Qin}, \bibinfo{person}{Yongyan Zheng}, \bibinfo{person}{Xipeng Qiu}, \bibinfo{person}{Xuanjing Huan}, {and} \bibinfo{person}{Tao Gui}.} \bibinfo{year}{2023}\natexlab{a}.
\newblock \showarticletitle{The Rise and Potential of Large Language Model Based Agents: {A} Survey}.
\newblock \bibinfo{journal}{\emph{CoRR}}  \bibinfo{volume}{abs/2309.07864} (\bibinfo{year}{2023}).
\newblock
\urldef\tempurl%
\url{https://doi.org/10.48550/ARXIV.2309.07864}
\showDOI{\tempurl}
\showeprint[arXiv]{2309.07864}


\bibitem[Xia et~al\mbox{.}(2023)]%
        {xia2023sheared}
\bibfield{author}{\bibinfo{person}{Mengzhou Xia}, \bibinfo{person}{Tianyu Gao}, \bibinfo{person}{Zhiyuan Zeng}, {and} \bibinfo{person}{Danqi Chen}.} \bibinfo{year}{2023}\natexlab{}.
\newblock \showarticletitle{Sheared llama: Accelerating language model pre-training via structured pruning}.
\newblock \bibinfo{journal}{\emph{arXiv preprint arXiv:2310.06694}} (\bibinfo{year}{2023}).
\newblock


\bibitem[Xiong et~al\mbox{.}(2023)]%
        {adversal2}
\bibfield{author}{\bibinfo{person}{Kai Xiong}, \bibinfo{person}{Xiao Ding}, \bibinfo{person}{Yixin Cao}, \bibinfo{person}{Ting Liu}, {and} \bibinfo{person}{Bing Qin}.} \bibinfo{year}{2023}\natexlab{}.
\newblock \showarticletitle{Examining the inter-consistency of large language models: An in-depth analysis via debate}. Association for Computational Linguistics.
\newblock


\bibitem[Yang et~al\mbox{.}(2023)]%
        {yang2023large}
\bibfield{author}{\bibinfo{person}{Zhengyi Yang}, \bibinfo{person}{Jiancan Wu}, \bibinfo{person}{Yanchen Luo}, \bibinfo{person}{Jizhi Zhang}, \bibinfo{person}{Yancheng Yuan}, \bibinfo{person}{An Zhang}, \bibinfo{person}{Xiang Wang}, {and} \bibinfo{person}{Xiangnan He}.} \bibinfo{year}{2023}\natexlab{}.
\newblock \showarticletitle{Large language model can interpret latent space of sequential recommender}.
\newblock \bibinfo{journal}{\emph{arXiv preprint arXiv:2310.20487}} (\bibinfo{year}{2023}).
\newblock


\bibitem[Yu et~al\mbox{.}(2023)]%
        {yu2023gptfuzzer}
\bibfield{author}{\bibinfo{person}{Jiahao Yu}, \bibinfo{person}{Xingwei Lin}, {and} \bibinfo{person}{Xinyu Xing}.} \bibinfo{year}{2023}\natexlab{}.
\newblock \showarticletitle{Gptfuzzer: Red teaming large language models with auto-generated jailbreak prompts}.
\newblock \bibinfo{journal}{\emph{arXiv preprint arXiv:2309.10253}} (\bibinfo{year}{2023}).
\newblock


\bibitem[Zhang et~al\mbox{.}(2023d)]%
        {agent4rec}
\bibfield{author}{\bibinfo{person}{An Zhang}, \bibinfo{person}{Leheng Sheng}, \bibinfo{person}{Yuxin Chen}, \bibinfo{person}{Hao Li}, \bibinfo{person}{Yang Deng}, \bibinfo{person}{Xiang Wang}, {and} \bibinfo{person}{Tat-Seng Chua}.} \bibinfo{year}{2023}\natexlab{d}.
\newblock \showarticletitle{On generative agents in recommendation}.
\newblock \bibinfo{journal}{\emph{arXiv preprint arXiv:2310.10108}} (\bibinfo{year}{2023}).
\newblock


\bibitem[Zhang et~al\mbox{.}(2023a)]%
        {fairness_1}
\bibfield{author}{\bibinfo{person}{Jizhi Zhang}, \bibinfo{person}{Keqin Bao}, \bibinfo{person}{Yang Zhang}, \bibinfo{person}{Wenjie Wang}, \bibinfo{person}{Fuli Feng}, {and} \bibinfo{person}{Xiangnan He}.} \bibinfo{year}{2023}\natexlab{a}.
\newblock \showarticletitle{Is chatgpt fair for recommendation? evaluating fairness in large language model recommendation}.
\newblock \bibinfo{journal}{\emph{arXiv preprint arXiv:2305.07609}} (\bibinfo{year}{2023}).
\newblock


\bibitem[Zhang et~al\mbox{.}(2023c)]%
        {agentcf}
\bibfield{author}{\bibinfo{person}{Junjie Zhang}, \bibinfo{person}{Yupeng Hou}, \bibinfo{person}{Ruobing Xie}, \bibinfo{person}{Wenqi Sun}, \bibinfo{person}{Julian McAuley}, \bibinfo{person}{Wayne~Xin Zhao}, \bibinfo{person}{Leyu Lin}, {and} \bibinfo{person}{Ji-Rong Wen}.} \bibinfo{year}{2023}\natexlab{c}.
\newblock \showarticletitle{Agentcf: Collaborative learning with autonomous language agents for recommender systems}.
\newblock \bibinfo{journal}{\emph{arXiv preprint arXiv:2310.09233}} (\bibinfo{year}{2023}).
\newblock


\bibitem[Zhang et~al\mbox{.}(2023e)]%
        {instructfollows}
\bibfield{author}{\bibinfo{person}{Junjie Zhang}, \bibinfo{person}{Ruobing Xie}, \bibinfo{person}{Yupeng Hou}, \bibinfo{person}{Wayne~Xin Zhao}, \bibinfo{person}{Leyu Lin}, {and} \bibinfo{person}{Ji-Rong Wen}.} \bibinfo{year}{2023}\natexlab{e}.
\newblock \showarticletitle{Recommendation as instruction following: A large language model empowered recommendation approach}.
\newblock \bibinfo{journal}{\emph{arXiv preprint arXiv:2305.07001}} (\bibinfo{year}{2023}).
\newblock


\bibitem[Zhang et~al\mbox{.}(2024)]%
        {knowledge_survey_2}
\bibfield{author}{\bibinfo{person}{Ningyu Zhang}, \bibinfo{person}{Yunzhi Yao}, \bibinfo{person}{Bozhong Tian}, \bibinfo{person}{Peng Wang}, \bibinfo{person}{Shumin Deng}, \bibinfo{person}{Mengru Wang}, \bibinfo{person}{Zekun Xi}, \bibinfo{person}{Shengyu Mao}, \bibinfo{person}{Jintian Zhang}, \bibinfo{person}{Yuansheng Ni}, {et~al\mbox{.}}} \bibinfo{year}{2024}\natexlab{}.
\newblock \showarticletitle{A Comprehensive Study of Knowledge Editing for Large Language Models}.
\newblock \bibinfo{journal}{\emph{arXiv preprint arXiv:2401.01286}} (\bibinfo{year}{2024}).
\newblock


\bibitem[Zhang et~al\mbox{.}(2023b)]%
        {collm}
\bibfield{author}{\bibinfo{person}{Yang Zhang}, \bibinfo{person}{Fuli Feng}, \bibinfo{person}{Jizhi Zhang}, \bibinfo{person}{Keqin Bao}, \bibinfo{person}{Qifan Wang}, {and} \bibinfo{person}{Xiangnan He}.} \bibinfo{year}{2023}\natexlab{b}.
\newblock \showarticletitle{Collm: Integrating collaborative embeddings into large language models for recommendation}.
\newblock \bibinfo{journal}{\emph{arXiv preprint arXiv:2310.19488}} (\bibinfo{year}{2023}).
\newblock


\bibitem[Zhao et~al\mbox{.}(2023)]%
        {zhao2023survey}
\bibfield{author}{\bibinfo{person}{Wayne~Xin Zhao}, \bibinfo{person}{Kun Zhou}, \bibinfo{person}{Junyi Li}, \bibinfo{person}{Tianyi Tang}, \bibinfo{person}{Xiaolei Wang}, \bibinfo{person}{Yupeng Hou}, \bibinfo{person}{Yingqian Min}, \bibinfo{person}{Beichen Zhang}, \bibinfo{person}{Junjie Zhang}, \bibinfo{person}{Zican Dong}, {et~al\mbox{.}}} \bibinfo{year}{2023}\natexlab{}.
\newblock \showarticletitle{A survey of large language models}.
\newblock \bibinfo{journal}{\emph{arXiv preprint arXiv:2303.18223}} (\bibinfo{year}{2023}).
\newblock


\bibitem[Zheng et~al\mbox{.}(2023)]%
        {zheng2023adapting}
\bibfield{author}{\bibinfo{person}{Bowen Zheng}, \bibinfo{person}{Yupeng Hou}, \bibinfo{person}{Hongyu Lu}, \bibinfo{person}{Yu Chen}, \bibinfo{person}{Wayne~Xin Zhao}, {and} \bibinfo{person}{Ji-Rong Wen}.} \bibinfo{year}{2023}\natexlab{}.
\newblock \showarticletitle{Adapting large language models by integrating collaborative semantics for recommendation}.
\newblock \bibinfo{journal}{\emph{arXiv preprint arXiv:2311.09049}} (\bibinfo{year}{2023}).
\newblock


\bibitem[Zhou et~al\mbox{.}(2023b)]%
        {lifeagent2}
\bibfield{author}{\bibinfo{person}{Shuyan Zhou}, \bibinfo{person}{Frank~F Xu}, \bibinfo{person}{Hao Zhu}, \bibinfo{person}{Xuhui Zhou}, \bibinfo{person}{Robert Lo}, \bibinfo{person}{Abishek Sridhar}, \bibinfo{person}{Xianyi Cheng}, \bibinfo{person}{Yonatan Bisk}, \bibinfo{person}{Daniel Fried}, \bibinfo{person}{Uri Alon}, {et~al\mbox{.}}} \bibinfo{year}{2023}\natexlab{b}.
\newblock \showarticletitle{Webarena: A realistic web environment for building autonomous agents}.
\newblock \bibinfo{journal}{\emph{arXiv preprint arXiv:2307.13854}} (\bibinfo{year}{2023}).
\newblock


\bibitem[Zhou et~al\mbox{.}(2023a)]%
        {zhou2023opportunities}
\bibfield{author}{\bibinfo{person}{You Zhou}, \bibinfo{person}{Xiujing Lin}, \bibinfo{person}{Xiang Zhang}, \bibinfo{person}{Maolin Wang}, \bibinfo{person}{Gangwei Jiang}, \bibinfo{person}{Huakang Lu}, \bibinfo{person}{Yupeng Wu}, \bibinfo{person}{Kai Zhang}, \bibinfo{person}{Zhe Yang}, \bibinfo{person}{Kehang Wang}, {et~al\mbox{.}}} \bibinfo{year}{2023}\natexlab{a}.
\newblock \showarticletitle{On the opportunities of green computing: A survey}.
\newblock \bibinfo{journal}{\emph{arXiv preprint arXiv:2311.00447}} (\bibinfo{year}{2023}).
\newblock


\end{thebibliography}

\appendix

\end{document}